\input xy \xyoption{all} \xyoption{arc}

\def\Tu#1#2#3{\save\POS,c="#2u";
p-(0,1)="v"**@{-}*{\bullet}*++!UC{#3},
"v"+(-#1,-#1)="vl";"vl"-(0,1)="#2l"**@{-},
"v"+(#1,-#1)="vr" ;"vr"-(0,1)="#2r"**@{-},
"vl";"vr",{\ellipse_{}} \restore }
\def\Tdr#1#2#3{\save\POS,c="#2r";
p-(0,1)="vr"**@{-},
"vr"+(-#1,-#1)="v"*{\bullet}*++!DC{#3};"v"-(0,1)="#2d"**@{-},
"v"+(-#1,#1)="vl" ;"vl"+(0,1)="#2l"**@{-},
"vl";"vr",{\ellipse^{}} \restore }
\def\Tdl#1#2#3{\save\POS,c="#2l";
p-(0,1)="vl"**@{-},
"vl"+(#1,-#1)="v"*{\bullet}*++!DC{#3};"v"-(0,1)="#2d"**@{-},
"v"+(#1,#1)="vr" ;"vr"+(0,1)="#2r"**@{-},
"vl";"vr",{\ellipse^{}} \restore }
\def\Cr#1#2{\save\POS,c="#2r";
p-(#1,0)-(#1,0)="#2l",{\ellipse^{}} \restore }
\def\Er#1#2{\save\POS,c="#2r";
p-(#1,0)-(#1,0)="#2l",{\ellipse_{}} \restore }
\def\Cl#1#2{\save\POS,c="#2l";
p+(#1,0)+(#1,0)="#2r",{\ellipse_{}} \restore }
\def\El#1#2{\save\POS,c="#2l";
p+(#1,0)+(#1,0)="#2r",{\ellipse^{}} \restore }

\def\alg      {algebra}
\def\assoc    {\varphi}
\def\be       {\begin{equation}}
\def\bearl    {\begin{array}{l}}
\def\bearll   {\begin{array}{ll}}
\def\Cat      {{\cal C}} 
\def\CC       {{\dl C}} 
\def\cft      {conformal field theory}
\def\cfts     {conformal field theories}
\def\coev     {c}   
\def\Cr       {\\ }
\def\deg      {{\rm deg}}
\def\dl       {\mathbb }
\def\ee       {\end{equation}}

\def\eear     {\end{array}}
\newcommand\erf[1] {(\ref{#1})}
\def\eq       {\,{=}\,}
\def\eval     {e}   
\def\fru      {{x}}  
\def\frv      {{y}}  
\def\H        {{\cal H}}
\def\Hat      {\breve}

\def\hrt      {\halfrootthree} 
\def\hy       {$\mbox{-\hspace{-.66 mm}-}$}
\def\I        {{\cal I}} 
\def\iatoprr  {{\scriptstyle r\in\I\atop \scriptstyle r=\hat r}}
\def\Im       {{\rm Im}}
\def\iN       {\,{\in}\,}
\def\irrep    {irreducible representation}
\newcommand\labl[1]{\label{#1}\ee}
\def\munit    {\varepsilon} 
\def\NB#1     {.... {\bf (NB} .... {\it #1} .... {\bf NB)} ....}
\def\nE       {\,{\not=}\,}
\newcommand\nxl[1] {\\{}\\[-.#1em]}
\def\ot       {\raisebox{.07em}{$\scriptstyle\otimes$}}
\def\oT       {\,\ot\,}
\def\qed      {\ \vrule height 5pt width 5pt depth 0pt}
\def\rep      {representation}
\def\rha      {rational Hopf algebra}
\newcommand\sfbox[2] {\mbox{\fbox{{\sc Figure} #2}}} 
\newcommand\sfpic[2] {\clearpage\fbox{{\sc Figure} #2}\\[1.5em]} 
\def\xtimes   {\times}     
\def\xTimes   {\Times}    
\def\SECTION  {\setcounter{equation}{0}\section}
\def\Times    {\,{\times}\,}
\def\tr       {{\rm tr}}
\def\ZZ       {{\dl Z}}

\catcode`\@=11
\def\citen#1{\if@filesw \immediate\write \@auxout {\string\citation{#1}}\fi%
\@tempcntb\m@ne \let\@h@ld\relax \def\@citea{}%
\@for \@citeb:=#1\do {\@ifundefined {b@\@citeb}%
    {\@h@ld\@citea\@tempcntb\m@ne{\bf ?}%
    \@warning {Citation `\@citeb ' on page \thepage \space undefined}}%
    {\@tempcnta\@tempcntb \advance\@tempcnta\@ne
    \setbox\z@\hbox\bgroup\ifcat0\csname b@\@citeb \endcsname \relax
    \egroup \@tempcntb\number\csname b@\@citeb \endcsname \relax
    \else \egroup \@tempcntb\m@ne \fi \ifnum\@tempcnta=\@tempcntb
    \ifx\@h@ld\relax \edef \@h@ld{\@citea\csname b@\@citeb\endcsname}%
    \else \edef\@h@ld{\hbox{--}\penalty\@highpenalty
    \csname b@\@citeb\endcsname}\fi
    \else \@h@ld\@citea\csname b@\@citeb \endcsname \let\@h@ld\relax \fi}%
\def\@citea{,\penalty\@highpenalty\hskip.13em plus.13em minus.13em}}\@h@ld}
\def\@citex[#1]#2{\@cite{\citen{#2}}{#1}}%
\def\@cite#1#2{\leavevmode\unskip\ifnum\lastpenalty=\z@\penalty\@highpenalty\fi%
  \ [{\multiply\@highpenalty 3 #1%
  \if@tempswa,\penalty\@highpenalty\ #2\fi}]}   %
\makeatother 
\catcode`\@=12

\documentclass[12pt]{article} \usepackage{amssymb,amsfonts}

\begin{document}


\begin{flushright}  {~} \\[-15 mm]  {\sf physics/9803038}\\[1mm]
{\sf March 1998} \end{flushright}
\vskip13mm

\centerline{\Large\bf $S_4$-symmetry of $6j$-symbols and Frobenius\hy Schur}
\vskip3mm
\centerline{\Large\bf indicators in rigid monoidal $C^*$-categories}
\vskip19mm

\centerline{\large J.\ Fuchs$\,^1_{}$,\, A.\ Ganchev$\,^2_{}$,\, 
K.\ Szlach\'anyi$\,^3_{}$\, and\, P.\ Vecserny\'es$\,^3_{}$}
\vskip13mm
\centerline{$^1_{}$ Max-Planck-Institut f\"ur Mathematik}
\centerline{Gottfried-Claren-Str.\ 26, \  D -- 53225~~Bonn, \ Germany}
\vskip2mm
\centerline{$^2_{}$ Institute for Nuclear Research and Nuclear Energy}
\centerline{72 Tsarigradsko Chaussee, \ BG -- 1784~~Sofia, \ Bulgaria}
\vskip2mm
\centerline{$^3_{}$ Central Research Institute for Physics}
\centerline{P.O.\ Box 49, \ H -- 1525~~Budapest \,114, \ Hungary}

\vskip5em
\begin{quote}{\bf Abstract}\\[1mm]
We show that a left-rigid monoidal $C^*$-category with irreducible monoidal
unit is also a sovereign and spherical category. Defining a Frobenius\hy Schur
type indicator we obtain selection rules for the fusion coefficients of 
irreducible objects. As a main result we prove $S_4$-invariance of 
$6j$-symbols in such a category. 
\end{quote}
\newpage 

\SECTION{Introduction} 

The quantities that are known as $6j$-{\em symbols\/} or 
$F$-{\em coefficients\/} 
appear in various disguises in mathematics and physics, for instance as:
recoupling coefficients in the theory of groups \cite{FAra,bibr,agbe} 
and quantum groups \cite{kire4}, as (partially gauge-fixed)
fusing matrices \cite{mose3,fefk3,liyu} in \cft,
as the Boltzmann weights for triangulations of three-manifolds
\cite{tuvi,bawe,TUra,kaMs} giving rise to topological lattice field theories
\cite{kaMs,dujn2,chfs}, as expansion coefficients of exchange algebra relations
in the algebraic theory of superselection sectors \cite{frrs2}, 
as the components of the 3-cocycle in Ocneanu's
non-Abelian cohomology \cite{ocneU},
as structure constants in the theory of 3-\alg s \cite{lawr2}, as specific 
endomorphisms of a von Neumann factor $M$ in the theory of subfactors 
\cite{evka5}, and as components
of the coassociator $\varphi$ of a \rha\ \cite{vecs,fugv3,fugv4}.
More generally, the $F$-coefficients can be described as the projections of the 
associativity constraint $\varphi$ of a semisimple, rigid, monoidal
category onto irreducible objects hence they map a pair of basic intertwiner
spaces to another pair. The $F$-coefficients depend on the choice of
basis (gauge choices) in the basic intertwiner spaces (spaces of 3-point
functions).
 
An important property of the $F$-coefficients is their
`tetrahedral symmetry', or more precisely, the fact that the gauge freedom
that is present in their definition can be fixed in such a way that they
are invariant under a set of transformations which form the permutation
group $S_4$. 
This symmetry property is for example needed in three-dimensional lattice
theories in order for the partition function that is defined in terms of the
$F$-coefficients to be independent of the triangulation so that the theory
is topological (see e.g.\ \cite{yett5,bawe}). Other places where this symmetry
plays an important r\^ole are the derivation of identities for `higher order
fusion coefficients' \cite{bave} and the construction of solutions to the
`big pentagon equation' \cite{bosz}.
Also, in pursuing a project for finding (possibly with the aid of computers)
explicit solutions of the Moore\hy Seiberg polynomial equations for `small'
fusion rings the assumption of tetrahedral symmetry allows for the substantial 
increase in the number of accessible fusion rings.
(In \cft\ the explicit computation of fusing matrices is e.g.\ required for
the calculation of the operator product coefficients.)

Let us remark that in \cft\ it seems to be common lore that the $F$-coefficients
possess $S_4$-invariance (see e.g.\ \cite{liyu,degi2}); but as 
a matter of fact no complete and detailed proof has ever been published.
In the theory of subfactors it was realized that the $S_4$-symmetry follows from
Frobenius reciprocity of intertwiners between bimodules \cite{evka5},
and in algebraic field theory the same aspects of the symmetry
are implicit in the results of \cite{frrs2} (see in particular the appendix
of that paper). The Frobenius maps between basic intertwiners generate 
the group $S_3$, and the coherent choice of bases in the orbits of this $S_3$ 
group lead to $S_4$-symmetric $F$-coefficients. In \cite{bawe} it was shown
that in order to have $S_4$-symmetry one does not need a braiding -- it is
sufficient that the category be spherical; however, in that paper the 
possibility of non-trivial Frobenius\hy Schur (FS) indicators 
\cite{ISaa,KIri,bant5} was not considered.

In this paper we present a proof that in a left-rigid monoidal $C^*$-category 
the $F$-coefficients possess $S_4$-symmetry. We would like to emphasize two 
fine points. The first is that, in general,  due to the possibility of having
non-trivial FS indicators, the Frobenius maps provide only a $\ZZ_2$-projective
representation of $S_3$; but nevertheless the corresponding signs cancel in the
transformation of the $F$-coefficients so that the $F$'s are truly 
$S_4$-invariant. The second
point is the following. The Frobenius transformations of order 3 map the
basic intertwiners space $(p\Times p, \hat p)$ into itself; hence its
eigenvalues are third roots of unity (here $\hat p$ is the conjugate of the 
irreducible object $p$). When considering $F$'s that involve such intertwiner 
spaces, then in order to verify
$S_4$-invariance we must calculate the different transforms of $F$
in different bases of the space $(p\Times p,\hat p)$. If, instead, we use only
a single
basis in this space, then the transforms of $F$ will possibly carry factors of
third roots of unity (this is illustrated on an example in the appendix).

Let us also briefly mention a possible application of our results to 
the quest of explicitly solving the polynomial equations for the braiding
and fusing matrices, which constitutes e.g.\ a part of the problem of
classifying all rational \cfts. Namely, one expects that when the category is 
modular, then the trace of the braiding matrices $R^{pp,q}$ can be expressed
completely in terms of the modular data, i.e.\ of the fusion rules, the
modular $S$ matrix and balancing phases (in the case of doubles 
of finite groups this was proved in \cite{bant5}).
{}Assuming that this expectation is indeed correct, it follows that 
the Frobenius\hy Schur indicators as well as the multiplicities of the 
above-mentioned third roots of unity do not constitute independent data, but 
are already determined uniquely by the modular data. In particular, one can 
immediately write down all the $F$-coefficients for which at least one line is
colored by the unit object as well as the $R$-coefficients. In fact, there 
are further special $F$-coefficients that can easily be solved for. 
Finally, our results concerning the $S_4$-symmetry allow to reduce
the number of unknowns in the polynomial equations drastically, namely for
generic $F$-coefficients by a factor of 24. 
\footnote{~Further simplifications 
come from the action of simple currents on the $F$-coefficients, which
should be derivable in a way similar to the $S_4$-action. Moreover, one would 
expect the pentagon equation itself, which can be regarded as the boundary
of a 4-simplex, to possess $S_5$-symmetry, just like the tetrahedra 
($F$-coefficients) -- the boundaries of 3-simplices -- possess $S_4$-symmetry.}

\SECTION{Rigid monoidal $C^*$-categories with irreducible mo\-no\-idal unit}

Our starting point is a left-rigid monoidal $C^*$-category
$(\Cat; (\munit, \xtimes, \{\lambda,\rho,\assoc\});
  (\ \widehat{}\,, \{\eval,\coev\}); {}^*)$
with the restriction that the monoidal unit $\munit$ is irreducible. Here 
$\xtimes$ is the monoidal product and $\lambda_a,\,\rho_a,\,\varphi_{a,b,c}$
with $a,b,c\in{\rm Obj}\,\Cat$ are natural isomorphisms
  \be  \bearl
  \lambda_a\colon\;\ a\to a\xtimes\munit\,,\qquad
  \rho_a\colon\;\ a\to \munit\xtimes a\,, \nxl6
  \varphi_{a,b,c}\colon\;\ a\xtimes (b\xtimes c)\to (a\xtimes b)\xtimes c
  \eear \ee
that satisfy the triangle identity 
  \be
  \varphi_{a,\munit,c}=(\lambda_a\xTimes 1_c)\,(1_a\xTimes\rho_c^{-1}),
  \labl{1.1a}
and the pentagon identity
  \be
  \varphi_{a\xtimes b,c,d}\,\varphi_{a,b,c\xtimes d}=(\varphi_{a,b,c}\xTimes
  1_d)\, \varphi_{a,b\xtimes c,d}\, (1_a\xTimes \varphi_{b,c,d})\,.\labl{1.1b}
The $C^*$-property requires the arrows (the intertwiners) 
  \be
  (a,b)\equiv {\rm Hom}\,(a,b):=\{ T\colon\ a\,{\to}\,b\}   \ee
between two objects 
$a,b$ to form a complex Banach space. The ${}^*$ is an involutive 
monoidal contravariant functor acting as identity on the objects and 
antilinearly on the intertwiner spaces. The norm of the intertwiners 
satisfies the $C^*$-property, $\Vert T^*T\Vert=\Vert T\Vert^2$, and the 
natural equivalences $\{\lambda,\rho,\assoc\}$ are isometries.
Irreducibility of $\munit$ implies that $(\munit,\munit)=\CC\cdot 1_\munit$.

The left conjugation $L\colon\;\Cat\to\Cat$ is an antimonoidal contravariant
(linear) functor, which is the identity on the monoidal unit. 
The functor $L$ is built up from a left rigidity structure $(\ \widehat{}
\,,\{\eval,\coev\})$, where $\ \widehat{}\ $ is the left conjugation 
on objects: $\hat a\equiv L(a)$. The left evaluation and coevaluation maps 
$\eval_a\colon\; \hat a \xtimes a \to \munit$ and
$\coev_a\colon\; \munit\to a\xtimes \hat a,\ a\iN{\rm Obj}\,\Cat$ satisfy
the left rigidity equations
  \be
  \lambda_a^*\,(1_a \xTimes \eval_a)\,\assoc_{a,\hat a, a}^*
  \,(\coev_a \xTimes 1_a)\,\rho_a = 1_a \,, \qquad
  \rho_{\hat a}^* \,(\eval_{a} \xTimes 1_{\hat a})
  \,\assoc_{\hat a, a, \hat a}\,(1_{\hat a}\xTimes\coev_a)
  \,\lambda_{\hat a} = 1_{\hat a}  \labl{1.2}
and lead to the definition 
  \be
  (a,b)\,\ni\, T \,\mapsto\, L(T) :=
  \rho_{\hat a}^* \,(\eval_b \xTimes 1_{\hat a})
  \,((1_{\hat b} \xTimes T ) \xTimes 1_{\hat a})
  \,\assoc_{\hat b,a,\hat a} \,(1_{\hat b} \xTimes \coev_a)
  \,\lambda_{\hat b} \,\in\, (\hat b,\hat a) \labl{1.3}
of the conjugated intertwiners $L(T)$.
Note that two left conjugations $L_1$ and $L_2$ that arise from left rigidity 
structures are always related by canonical natural equivalences 
$\mu\colon\ L_1\to L_2$ given by
  \be \bearll
  \mu_a\!\!\!&=\rho_{a^2}^*\,(e_a^1\xtimes 1_{a^2})\,\assoc_{a^1,a,a^2}\,
  (1_{a^1}\xtimes c_a^2)\,\lambda_{a^1}\in(a^1,a^2)\,, \nxl8
  \mu_a^{-1}\!\!\!&=\rho_{a^1}^*\,(e_a^2\xtimes 1_{a^1})\,\assoc_{a^2,a,a^1}\,
  (1_{a^2}\xtimes c_a^1)\,\lambda_{a^2}\in(a^2,a^1)\,, \eear
  \qquad \mbox{for every }a\iN{\rm Obj}\,\Cat\,,  \labl{1.4}
where $a^i\equiv L_i(a)$ for $i=1,2$. 

The $C^*$-property allows us to define a right rigidity structure within $\Cat$:
  \be
  (R(a),e^R_a,c^R_a):=(L(a),c^{L*}_a,e^{L*}_a)\equiv (\hat a,c^*_a,e^*_a)
  \quad {\rm for}\;\ a\iN{\rm Obj}\,\Cat\,,  \labl{1.6a}
leading to a right conjugate functor $R\colon\;\Cat\to\Cat$ 
similarly to \erf{1.3}: 
  \be \bearll
  (a,b)\,\ni\, T \,\mapsto\, R(T) \!\!&:=
  \lambda_{R(a)}^*\,(1_{R(a)}^{}\xTimes\eval_b^R)\,
  (1_{R(a)}^{} \xTimes (T\xTimes 1_{R(b)}^{}))\,
  \assoc_{R(a),a,R(b)}^*\, (\coev_a^R\xTimes 1_{R(b)}^{})\,
  \rho^{}_{R(b)}\nxl8 &\,\equiv
  \lambda_{\hat a}^*\,(1_{\hat a}\xTimes\coev_b^*)\,
  (1_{\hat a}^{} \xTimes (T\xTimes 1_{\hat b}^{}))\,
  \assoc_{\hat a,a,\hat b}^*\, (\eval_a^*\xTimes 1_{\hat b}^{})
  \rho_{\hat b}^{}\,\in\, (\hat b , \hat a)\, .  \eear \labl{1.6b}
Since the right and left conjugated objects are identical, 
$R(a)=L(a)\equiv\hat a$ for all $a\iN{\rm Obj}\,\Cat$, the 
canonical natural isomorphisms
  \be
  \{\kappa_a\colon\ R(L(a))\to a\}\,,\qquad 
  \{\tilde\kappa_a\colon\ L(R(a))\to a\}\,, \labl{1.7} 
are given by
  \be \bearl
  \kappa_a=\lambda_a^*\,(1_a\xTimes e^R_{L(a)})\,\varphi^*_{a,L(a),R(L(a))}\,
                (c_a^L\xTimes 1_{R(L(a))})\,\rho_{R(L(a))}^{}
    \equiv \lambda_a^*\,(1_a\xTimes c^*_{\hat a})\,\varphi^*_{a,\hat a,\hat
           {\hat a}}\,(c_a\xTimes 1_{\hat{\hat a}})\,\rho_{\hat{\hat a}}\,,
  \nxl8
  \tilde\kappa_a=\rho_a^*\,(e^L_{R(a)}\xTimes 1_a)\,\varphi_{L(R(a)),R(a),a}^{}
                \,(1_{L(R(a))}\xTimes c_a^R)\,\lambda_{L(R(a))}
    \equiv \rho_a^*\,(e_{\hat a}\xTimes 1_a)\,\varphi_{\hat{\hat a},\hat a,a}\,
                (1_{\hat{\hat a}}\xTimes e_a^*)\,\lambda_{\hat{\hat a}} \,, 
  \eear \labl{1.8}
which satisfy
  \be \bearll 
  \kappa_a^{-1}= \lambda_{\hat{\hat a}}^*\,(1_{\hat{\hat a}}\xTimes e_a)\,
            \varphi^*_{\hat{\hat a},\hat a,a}\,
                (e_{\hat a}^*\xTimes 1_a)\,\rho_a=\tilde\kappa^*_a\,,\nxl8
  \tilde\kappa_a^{-1}= \rho_{\hat{\hat a}}^*\,(c_a^*\xTimes 1_{\hat{\hat a}})\,
            \varphi_{a,\hat a,\hat{\hat a}}\,
                (1_a\xTimes c_{\hat a})\,\lambda_a=\kappa^*_a\,. \eear
  \labl{1.9}
One can define (positive) left and right inverses for 
$a\iN{\rm Obj}\,\Cat$ \cite{loro}: 
  \be
  \Phi^L_a\colon\ (a\xTimes b,a\xTimes c)\to (b,c)\,,\quad\ 
  \Phi^R_a\colon\ (b\xTimes a,c\xTimes a)\to (b,c)\,,
  \quad {\rm for}\ b,c\iN{\rm Obj}\,\Cat
  \ee
using evaluation and coevaluation maps. In the special case $b=c=\munit$ 
these maps
  \be \bearl
  (a,a)\,\ni\, T\mapsto \Phi^L_a(T):=e_a\,(1_{\hat a}\xTimes T)\,e_a^*
       \,\in (\munit,\munit)=\CC\cdot 1_\munit \nxl8
  (a,a)\,\ni\, T\mapsto \Phi^R_a(T):=c_a^*\,(T\xTimes 1_{\hat a})\,c_a
       \,\in (\munit,\munit)=\CC\cdot 1_\munit \eear \labl{1.10}       
are faithful positive linear functionals. Since they are bounded from
below \cite{loro},
  \be \bearl
  (a,a)\ni\, T^*T\leq\rho_a^*\,(c_a^*c_a\xTimes 1_a)\rho_a\cdot
  \lambda_a^*\,(1_a\xTimes\Phi^L_a(T^*T))\,\lambda_a\,, \nxl8
  (a,a)\ni\, T^*T\leq\lambda_a^*\,(1_a\xTimes e_ae_a^*)\lambda_a\cdot
  \rho_a^*\,(\Phi^R_a(T^*T)\xTimes 1_a)\,\rho_a\,, \eear \qquad{\rm for}\
  T\iN(a,b)\,, \labl{1.11}
it follows that ${\rm End}\, a\equiv (a,a)$ is finite-dimensional for every
$a\iN{\rm Obj}\,\Cat$; therefore $\Cat$ is semisimple.
We denote the subset of irreducible objects by $\I\subset{\rm Obj}\,\Cat$.

Semisimplicity allows us to construct the so-called {\em standard\/} rigidity 
intertwiners \cite{loro}, leading to a left conjugation functor $L_s$ among 
the equivalent ones (cf.\ \erf{1.4}) that obeys $L_s=R_s$ by \erf{1.6a} and
\erf{1.6b}. This conjugation is achieved by the following procedure.

In the case of an irreducible object $p\iN\I$ one 
uses the scalar freedom $e_p\mapsto z_pe_p,\; c_p\mapsto z_p^{-1}c_p$ with
$z_p\iN\CC\,{\setminus}\{0\}$ to set
  \be
  e_p^{} e_p^*=d_p\,1_\munit=c_p^* c_p^{}\qquad
  \mbox{for every }p\iN\I\,,\labl{1.12}
where $d_a$ is the {\em quantum\/} or {\em statistics dimension\/} of 
an arbitrary object $a\iN\Cat$, defined to be
  \be
  d_a:=\Vert e_a\Vert\,\Vert c_a\Vert\,.\labl{1.5}

Then for an arbitrary object $a$ one chooses two orthogonal and complete 
sets of partial isometries
  \be  \bearll
  V_a^{p\alpha}\colon& p\to a\,,\nxl8 W_a^{p\alpha}\colon& \hat p\to \hat a\,,
  \eear \qquad {\rm for}\ p\iN\I,\; \alpha\eq1,\dots, m_a^p\,, \labl{1.13a}
satisfying
  \be
  V_a^{p\alpha*} V_a^{p'\alpha'}=\delta_{pp'}\delta_{\alpha\alpha'}1_p\,,\qquad
  W_a^{p\alpha*} W_a^{p'\alpha'}=\delta_{pp'}\delta_{\alpha\alpha'}1_{\hat p}\,,
  \labl{1.13b}
and
  \be
  \sum_{p,\alpha}V_a^{p\alpha} V_a^{p\alpha*}=1_a\,,\qquad
  \sum_{p,\alpha}W_a^{p\alpha} W_a^{p\alpha*}=1_{\hat a}\,,
  \labl{1.13c}
to define
  \be
  e_a:=\sum_{p,\alpha}e_p(W_a^{p\alpha*}\xTimes V_a^{p\alpha*})\in
      (\hat a\Times a,\munit)\,,\quad\
  c_a:=\sum_{p,\alpha}(V_a^{p\alpha}\xTimes W_a^{p\alpha})c_p\in
      (\munit,a\Times \hat a)\,.\labl{1.14}
Then $(\ \widehat{}\;,\{ e_a,c_a\})$ becomes a left rigidity 
structure, i.e.\ it satisfies \erf{1.2} due to naturality of 
$\{\lambda,\rho,\varphi\}$ and \erf{1.13b}, \erf{1.13c}. The rigidity 
intertwiners $(e_a, c_a)$ of an object $a$ satisfying \erf{1.12} and \erf{1.14} 
are called {\em standard\/}. 

Now the corresponding right rigidity structure (cf.\ \erf{1.6a} and \erf{1.6b}) 
$(\ \widehat{}\;,\{ c_a^*,e_a^*\})$ leads to a right conjugation
functor $R$ that is identical to the left conjugation $L$. As a matter 
of fact, by an explicit calculation one obtains
  \be
  (b,a)\,\ni\, T\,\mapsto\, R(T)= L(T)=\sum_{p,\alpha,\beta}
  W_b^{p\beta}t_p^{\alpha\beta}1_{\hat p}W_a^{p\alpha*}\,\in\,(\hat a,\hat b),
  \labl{1.15a}
with
  \be
  t_p^{\alpha\beta}1_p:=V_a^{p\alpha*}TV_b^{p\beta},\quad 
  t_p^{\alpha\beta}\iN\CC\,.\labl{1.15b}
Therefore in case of standard conjugation we use the notation 
$\hat T\equiv R(T)=L(T)$ for the conjugated intertwiners. 

One can also prove \cite{loro} that if $(e_a,c_a)$ and $(e_a',c_a')$ are both 
standard, then
  \be
  e_a'=e_a(1_{\hat a}\xTimes U_a),\qquad c_a'=(U_a^*\xTimes 1_{\hat a})c_a,
  \labl{1.16}
where $U_a\iN (a,a)$ is unitary. Moreover, 
  \be \bearll
  e_{(a,b)}:=
  e_b\,(1_{\hat b}\xTimes\rho_b)\,(1_{\hat b}\xTimes (e_a\xTimes 1_b))\,
  (1_{\hat b}\xTimes\varphi_{\hat a,a,b})\,\varphi_{\hat b,\hat a,a\xtimes b}^*
  & \in((\hat b\xTimes\hat a)\xTimes (a\xTimes b),\munit) \nxl8
  c_{(a,b)}:=\varphi_{a\xtimes b,\hat b,\hat a}^*\,
              (\varphi_{a,b,\hat b}\xTimes 1_{\hat a})\,
              ((1_a\xTimes c_b)\xTimes 1_{\hat a})\,
              (\lambda_a\xTimes 1_{\hat a})\,c_a
  & \in(\munit,(a\xTimes b)\xTimes (\hat b\xTimes\hat a)) \eear \labl{1.17}
are standard if $(e_a,c_a)$ and $(e_b,c_b)$ are standard. Thus in case of
standard conjugation, $L=R$, the natural equivalence $\{\alpha_{a,b}\colon\;
\widehat{a{\xtimes}b}\to \hat b\xTimes \hat a\}$ that expresses the 
antimonoidality of the conjugation functor is given by
  \be \bearll
  \alpha_{a,b}=\rho_{\hat b\xtimes\hat a}^*(e_{a\xtimes b}^{}\xTimes 
  1_{\hat b\xtimes\hat a})\,
  \varphi_{\widehat{a\xtimes b},a\xtimes b,\hat b\xtimes\hat a}
  (1_{\widehat{a\xtimes b}}\xTimes c_{(a,b)}^{})\,\lambda_{\widehat{a\xtimes b}}
  & \in(\widehat{a\xTimes b},\hat b\xTimes\hat a)\,, \nxl8
  \alpha_{a,b}^{-1}=\rho_{\widehat{a\xtimes b}}^*(e_{(a,b)}^{}\xTimes 
  1_{\widehat{a\xtimes b}})\,
  \varphi_{\hat b\xtimes\hat a,a\xtimes b,\widehat{a\xtimes b}}\,
  (1_{\hat b\xtimes \hat a}\xTimes c_{a\xtimes b}^{})\,
  \lambda_{\hat b\xtimes \hat a}
  &\in(\hat b\xtimes\hat a,\widehat{a\xtimes b})\,.  \eear \labl{1.18}

\SECTION{Sovereignty, traces and sphericity}

We note that in case of standard rigidity intertwiners the choice 
$\psi_a:=1_{\hat a}$ for every $a\iN{\rm Obj}\,\Cat$ leads to a natural 
equivalence $\psi\colon\; R\to L$ and makes $\Cat$ a {\it sovereign\/} 
category \cite{frye3}. Indeed, $\psi_\munit=1_\munit$ and monoidality of $\psi$ 
is clear, one has only to show the commutativity of the sovereignty diagram, 
which reads in this case as $\kappa_a\eq\tilde\kappa_a$. 
Since $\kappa$ and $\tilde\kappa$ are natural
equivalences it is enough to show this equality for $p\iN\I$. But then one has
  \be
  \kappa_p^*\kappa_p^{}=x_p^{}1_{\hat{\hat p}} \quad{\rm and}\quad
  \tilde\kappa_p^*\tilde\kappa_p^{}=\tilde x_p^{}1_{\hat{\hat p}}
  \quad\mbox{for all }\; p\iN\I\,, \labl{2.1}
with $x_p,\tilde x_p\in{\dl R}_+$ and $x_p=\tilde x_p^{-1}$ due to \erf{1.9}.
Using \erf{1.12} and rigidity one obtains
  \be \bearl
  x_pd_{\hat p}\,1_\munit
   =e_{\hat p}\,(x_p1_{\hat{\hat p}}\xTimes 1_{\hat p})\,e_{\hat p}^*
   =e_{\hat p}\,(\kappa_p^*\kappa_p\xTimes 1_{\hat p})\,e_{\hat p}^*
   =c_p^*c_p^{}=d_p\,1_\munit \,, \nxl6
  \tilde x_pd_{\hat p}\,1_\munit
   =c_{\hat p}^*\,(1_{\hat p}\xTimes \tilde x_p1_{\hat{\hat p}})\,c_{\hat p}
   =c_{\hat p}^*\,(1_{\hat p}\xTimes\tilde\kappa_p^*\tilde\kappa_p)\,c_{\hat p}
   =e_p^{}e_p^*=d_p\,1_\munit\,,  \eear \labl{2.2}
which imply $x_p=\tilde x_p$, hence $x_p=\tilde x_p=1$ and $d_{\hat p}=d_p$.
Then \erf{2.1} and \erf{1.9} lead to the equalities $\tilde \kappa_p=\kappa_p,\;
p\iN\I$, of isometries. Owing to naturality the maps
$\tilde\kappa_a\equiv\kappa_a\colon\;
\hat{\hat a}\to a,\; a\iN{\rm Obj}\,\Cat$, are isometries.

Having reached a standard conjugation, $L\eq R$, and sovereignty, let us 
consider the conjugates of standard rigidity intertwiners. 
Using \erf{1.18} one obtains
  \be
  \alpha_{\hat a,a}\hat e_a=c_{\hat a}
       \in (\munit,\hat a\xTimes\hat{\hat a})\,,\qquad
  \hat c_a\alpha_{a,\hat a}^{-1}=e_{\hat a}
       \in (\hat{\hat a}\xTimes\hat a,\munit)\,,\labl{2.3a}
and the rigidity intertwiners 
of conjugated objects are related to the original ones as
  \be
  e_{\hat a}=c_a^*\,(\kappa_a^{}\xTimes 1_{\hat a})\,,\qquad
  c_{\hat a}=(1_{\hat a}\xTimes\kappa_a^{-1})\,e_a^*\,,\labl{2.3b}
as is seen by using \erf{1.8} and \erf{1.9}. 
{}From \erf{2.3b} it follows that 
further specification of the rigidity intertwiners, e.g.\ identification of the 
$(e_a,c_a)$ and $(c_{\hat a}^*,e_{\hat a}^*)$ rigidity pairs, may be achieved
only in case of involutive conjugation, which will be discussed in Chapter 
\ref{C4}.

In a sovereign monoidal category one can define left and right traces ${\rm tr}
^{L/R}_a\colon\; (a,a)\to(\munit,\munit)$ for each $a\iN{\rm Obj}\,\Cat$:
  \be \bearl
  \tr_a^L(T):=e_a^L\,(\psi_a^{}\xTimes T)\,c_a^R\,, \nxl7
  \tr_a^R(T):=e_a^R\,(T\xTimes \psi_a^{})\,c_a^L\,, \eear \qquad {\rm for}\;\
  \psi_a\colon\ R(a)\to L(a)\,, \labl{2.4}
obeying the property   
  \be
  \tr^{R/L}_a(TS)=\tr^{R/L}_b(ST)\qquad\mbox{for all}\;\ T\iN(b,a),\ S\iN(a,b)
  \,. \labl{2.5}
In the case of standard conjugation in a $C^*$-category, together with the 
previously chosen sovereignty maps, $\{\psi_a=1_{\hat a}\}$, these left
(right) traces become identical to left (right) inverses \erf{1.10}, hence
they are positive traces. Moreover, due to standardness they are equal:
  \be
  \tr_a^L(T)=\tr_a^R(T)=:\tr_a(T)\qquad\mbox{for all}\;\ T\iN(a,a)\,; \labl{2.6}
hence $\Cat$ is also a {\it spherical\/} category \cite{bawe}.

\SECTION{Frobenius\hy Schur indicators}

For any finite group $G$ one defines the Frobenius\hy Schur element $\sigma$
of the group algebra $\CC G$ as in \cite{ISaa}, i.e.\ (up to normalization) 
  \be
  \sigma:={1\over\vert G\vert}\sum_{g\in G}g^2\,.\labl{3.1}
The Frobenius\hy Schur element is a central selfadjoint element of $\CC G$; its
central decomposition reads
  \be
  \sigma=\sum_{r\in\I}\frac{\nu_r}{d_r}\,e_r\qquad {\rm with}\qquad
  \nu_r=\left\{\begin{array}{rl} 0 &{\rm for}\,\ r\nE\hat r\,,\\
                       \pm 1&{\rm for}\,\ r\eq\hat r\,, \eear \right. \labl{3.2}
where $e_r,\;r\iN\I$, are the minimal central projectors and 
$d_r\eq\tr_r(e_r)$ is the (integral) dimension of the corresponding
simple ideal of the group algebra $\CC G$. The three-valent indicator $\nu_r$ 
that is defined by \erf{3.2}
is the {\em Frobenius\hy Schur indicator\/}, which is zero 
on non-selfconjugate simple ideals while on selfconjugate simple ideals the 
$+(-)$ sign indicates (pseudo)reality. 

There is an extension of $\sigma$ for $C^*$-Hopf \cite{SWee} or weak 
$C^*$-Hopf algebras \cite{bosz}, which is obtained  using the Haar integral 
$h\iN H$ of the (weak) $C^*$-Hopf algebra $H$:
  \be
  \sigma:={1\over\epsilon({\bf 1})} h^{(1)}\,h^{(2)} \,;  \labl{3.3}
here we use Sweedler notation for the coproduct, i.e.\
$h^{(1)}\oT h^{(2)}\equiv\Delta(h)$,
and ${\bf 1}$ and $\epsilon$ are the unit and the counit of $H$, 
respectively. Since weak $C^*$-Hopf algebras contain $C^*$-Hopf algebras as 
special cases, we prove the property \erf{3.2} for $\sigma$ as defined
in \erf{3.3} only for the former case.

There is a unique positive $g\iN H$ \cite{szla3} that implements the square of 
the antipode, i.e.\ $gxg^{-1}\eq S^2(x)$ for all $x\iN H$, and satisfies the 
normalization condition\,
${\rm tr}\, D_r(g)={\rm tr}\, D_r(g^{-1})=:\tau_r\iN{\dl R}_+$. 
Then $S(g)\eq g^{-1}$ holds, and if the unit representation is irreducible 
then $\tau_r=d_r/\epsilon({\bf 1})$. Using the properties \cite{bons}
  \be
  h^2=h^*=S(h)=h\,,\qquad  
  h^{(1)}\oT xh^{(2)}y=S(x)h^{(1)}S^{-1}(y)\oT h^{(2)}\quad\mbox{for all }
  x,y\iN H \labl{3.4a}
and
  \be
  h^{(2)}\oT h^{(1)}=h^{(1)}\oT gh^{(2)}g\,,\qquad
  S(h^{(1)})\oT h^{(2)}=\sum_{r\in\I}{1\over\tau_r}\sum_{i,j}
  e_r^{ij}g^{-1/2}\oT g^{-1/2}e_r^{ji} \labl{3.4b}
of the Haar integral $h\iN H$, one obtains
  \be
  \sigma^*={1\over\epsilon({\bf 1})}\, h^{(2)}h^{(1)}= 
  {1\over\epsilon({\bf 1})}\,h^{(1)}gh^{(2)}g=
  {1\over\epsilon({\bf 1})}\,h^{(1)}S^{-1}(g)gh^{(2)}=\sigma \labl{3.5}
and
  \be\bearll
  \sigma\cdot y \!\!
  &= \displaystyle\frac1{\epsilon({\bf 1})}\, h^{(1')}h^{(1)}h^{(2')}h^{(2)}
     \cdot y= \displaystyle\frac1{\epsilon({\bf 1})}\, S(h^{(1)})h^{(1')}
     h^{(2')}h^{(2)}\cdot y\nxl8
  &= \displaystyle\frac1{\epsilon({\bf 1})}\, S(h^{(1)}S^{-1}(y))h^{(1')}
     h^{(2')}h^{(2)} =y\cdot \sigma \eear \labl{3.6}
for all $y\iN H$,
i.e.\ $\sigma$ is a central selfadjoint element of $H$. From the second identity
in \erf{3.4b} it follows that $\sigma$ vanishes on non-selfconjugate ideals, 
hence $\nu_r\eq0$ for $r\nE\hat r$. For a selfconjugate ideal, $e_r\eq 
e_{\hat r}$, one obtains  
  \be \bearll
  \epsilon({\bf 1})\cdot e_r\sigma \!\!
  &= \displaystyle\frac1{\tau_r}\sum_{i,j}S^{-1}(e_r^{ij}g^{-1/2})
  g^{-1/2}e_r^{ji} \nxl8
  &= \displaystyle\frac1{\tau_r}\sum_{i,j}g^{1/2}S^{-1}(e_r^{ij})
  g^{-1/2}e_r^{ji}= \displaystyle\frac1{\tau_r}\sum_{i,j}S_0(e_r^{ij})e_r^{ji}
  \,, \eear \labl{3.7}
where $S_0:={\rm Ad}_{g^{1/2}}\,{\circ}\,S^{-1}$ is the involutive antipode: 
$S_0^2={id}_H,\ S_0\circ{}^*={}^*\circ S_0$. Since $e_r=e_{\hat r}$ for matrix 
units, it follows that
$S_0(e_r^{ij})=v_re_r^{ji}v_r^*$ with $v_r\iN e_rH$ unitary. Moreover,
  \be
  e_r^{ij}=S_0^2(e_r^{ij})=S_0(v_r^*)v_r^{} e_r^{ij}v_r^*S_0(v_r)\labl{3.8}
implies that $S_0(v_r^*)v_r$ is a central unitary element in $e_rH$, hence
$S_0(v_r^*)=\pm v_r^*=:\nu_rv_r^*$ for $r=\hat r$ due to the involutivity
of $S_0$. But then  
  \be \bearll \sigma \!\!
  &=\displaystyle\sum_\iatoprr
    \frac1{\epsilon({\bf 1})\tau_r}\sum_{i,j} S_0(e_r^{ij})e_r^{ji}
   =\displaystyle\sum_\iatoprr\frac1{d_r}\sum_{i,j} v_r^{}e_r^{ji}v_r^*e_r^{ji}
  \nxl8
  &=\displaystyle\sum_\iatoprr\frac1{d_r}\,v_r^{}v_r^{*T} 
   =\displaystyle\sum_\iatoprr\frac1{d_r}\,v_r^{}v_r^*S_0(v_r^*)v_r^{}
   =\sum_{r\in\I}\frac{\nu_r}{d_r}\,e_r \,, \eear \labl{3.9}
proving \erf{3.2}.

There exists a purely categorical definition of the Frobenius\hy Schur
indicator. Let $\Cat$ be a category as in section 2 and $p$ an 
irreducible object. If $p$ is selfconjugate, 
then there exists an invertible intertwiner $J_p\colon\;
p\to \hat p$. Let us define the self-intertwiner
  \be
  \nu_p:=J_p^{-1}\hat J_p^{}\kappa_p^{-1}\in(p,p)\,,\labl{3.10}
which is independent of the choice of $J_p$ due to irreducibility of $p$
and linearity of the conjugation. Hence $\nu_p$ is an isometry, because an 
isometric $J_p$ can be chosen. In order to prove that even 
$\nu_p=\pm 1_p$ holds, one first computes that
  \be
  \Phi_p^R(\nu_p):=c_p^*\,(\nu_p\xTimes 1_{\hat p})\,c_p=
  c_p^*\,(J_p^{-1}\xTimes J_p)\,e_p^*\,,\labl{3.11}
using \erf{1.3} and \erf{1.9}. However, the rigidity intertwiners
  \be
  (e_p',c_p'):=(c_p^*(J_p^{-1}\xTimes J_p),(J_p^{-1}\xTimes J_p)e_p^*) \,\in
  ((\hat p\xTimes p,\munit),(\munit,p\xTimes\hat p))\labl{3.12}
are standard, hence owing to \erf{1.16} we have
  \be
  e_p'=e_pu_p\quad{\rm and}\quad c_p'=u_p^*c_p\quad\mbox{for some }\; u_p\iN\CC\
  \;{\rm with}\;\ \vert u_p\vert\eq1 \,. \labl{3.13}
Therefore
  \be
  u_p^{}d_p^{}=u_p^{}e_p^{}e_p^*=e_p'e_p^*=\Phi_p^R(\nu_p^{})
  =c_p^*c_p'=u_p^*c_p^*c_p^{}= u_p^*d_p^{}\,,\labl{3.14}
which proves that $u_p=\pm 1$ and $\nu_p=\pm 1_p$. 

Defining $\nu_p$ as the zero intertwiner for non-selfconjugate irreducible 
objects $p$, one can obtain a natural map $\nu$ between the identity functors. 

Since the representation category of a (pure weak) $C^*$-Hopf algebra is 
a rigid monoidal $C^*$-category with irreducible unit object,
the consistency of the Hopf algebraic and the categorical definitions 
of the FS indicator requires that $\nu_p\eq d_pD_p(\sigma)$ for $p\iN\I$, 
where $D_p\colon\; H\to {\rm End}\,V_p$\, is an \irrep\ of $H$.
This is indeed the case: One can
use the natural equivalences $\lambda,\rho$, the standard rigidity 
intertwiners that were given in \cite{bosz}, and the involutive conjugation
on the objects $D\mapsto \hat D:=D^T\circ S_0$ to arrive 
at $\kappa_D=1_D$. Then for an irreducible selfconjugate object $D_p$ 
the choice $J_p^{\hat\alpha\alpha}:=D_p^{\hat\alpha\alpha}(S_0(v_p^*))$
for the matrix elements of the map $J_p\colon\; V_p\to\hat V_p$ leads to the 
desired result. 

\SECTION{Involutive conjugation}\label{C4}

The existence of the natural isomorphism $\{\kappa_a\colon\;\hat{\hat a}\to a\}$
suggests that a convenient choice $\hat{\hat a}=a$ can be achieved. 
This requires, however, a further assumption on the category $\Cat$; namely, the
cardinalities of the sets of objects in any two equivalence classes that are
related by conjugation must be 
equal. This quite harmless assumption will always hold after adjoining, if
necessary, new objects to $\Cat$, and this procedure does not change the 
category within equivalence.

Therefore from now on
we assume that the object map $a\mapsto \hat a$ of our conjugation
functor is involutive, $\hat{\hat a}\eq a$ for all $a\in{\rm Obj}\,\Cat$, and
that $\hat a\eq a$ whenever $\hat a\simeq a$. It follows that every choice of
standard rigidity intertwiners leads to a conjugation functor $L\equiv R$ 
that is involutive on the arrows as well. As a matter of fact, for each
$T\iN(a,b)$ we have $T^{LL}=T^{LR}=\kappa_b^{-1}T\kappa_a=T$.

Due to the involutivity of the conjugation, for irreducible objects $p$ the
natural isomorphism $\kappa_p\colon\;\hat{\hat p}=p\to p$ becomes 
$\kappa_p\eq\chi_p1_p$ with $\chi_p\iN\CC$ and $\vert\chi_p\vert\eq1$. The 
question arises whether the scalar $\chi_p$ that appears here has something to 
do with the Frobenius\hy Schur indicator or it can be ``gauged'' away. 
The relations \erf{2.3b} now read as
  \be \bearl
  e_{\hat p}=c_p^*\,(\kappa_p^{}\xtimes 1_{\hat p})=\chi_p^{}\,c_p^* \,, \nxl8
  c_{\hat p}=(1_{\hat p}\xtimes\kappa_p^{-1})\,e_p^*=\chi_p^{-1}e_p^*
  \equiv\bar\chi_p^{}\,e_p^*\,. \eear \labl{4.1}
Inserting $p\eq\hat q$ in the first equation, we obtain 
$\chi_{\hat q}e_{\hat{\hat q}}^*\eq c_{\hat q}\eq\bar\chi_qe_q^*$, implying
that $\chi_{\hat q}=\bar\chi_q\equiv\chi_q^{-1}$.
For selfconjugate irreducibles $q\eq\hat q$ we obtain $\chi_q=\pm 1$. In this 
case by choosing the map $J_q\colon\; q\to\hat q$ in the definition \erf{3.11}
of the FS indicator to be $1_q$, we obtain 
  \be
  \nu_q^{}=J^{-1}\hat J\kappa_q^{-1}=1_q1_q(\chi_q^{-1}1_q)=\chi_q^{-1}1_q 
  \qquad\mbox{for all }\; q\iN\I\;\ {\rm with}\;\ q\eq\hat q \,.\labl{4.2}
If $p\not=\hat p$ for $p\iN\I$, then we can employ the freedom \erf{1.16} so as
to change the rigidity intertwiners of one of the irreducible objects of the 
pair $\{ p,\hat p\}$, let us say of $\hat p$:
  \be
  e_{\hat p}':=\chi_p^{-1}e_{\hat p}\,,\quad c_{\hat p}':=\chi_pc_{\hat p}\,,
  \qquad e_p':=e_p,\quad c_p':=c_p\,.\labl{4.3}
Then $e_{\hat p}'=c_p^{\prime *}$ and $c_{\hat p}'=e_p^{\prime *}$, i.e.
the coefficients $\chi$ that are present in \erf{4.1} are gauged away for 
$p\iN\I$, $p\nE\hat p$.

To summarize: after enlarging $\Cat$ appropriately, there exists a {\em standard
conjugation functor\/} that is involutive both 
on the objects and on the arrows and has the property that for $p\iN\I$
  \be \bearl
  e_{\hat p}=\chi_p\,c_p^*\,, \nxl9
  c_{\hat p}=\chi_p^{-1}e_p^*\,, \eear \qquad{\rm with}\quad
  \chi_p=\left\{\bearll 1 & {\rm for}\,\ p\nE\hat p\,, \nxl9
                     \nu_p & {\rm for}\,\ p\eq\hat p\,. \eear \right. \labl{4.4}

\SECTION{Frobenius maps on basic intertwiner spaces}\label{fmbi}

Let us call the space $(p,q\xTimes r)$, for $p,q,r\iN\I$,
a {\em basic intertwiner space\/}. It is a Hilbert space with
scalar product determined by
  \be
  (t_1,t_2)\,1_p := t^*_1 t_2 \qquad{\rm for}\;\ t_1,t_2 \iN (p,q\xTimes r)
  \,. \labl{5.1}
We define two antilinear maps 
  \be
  \fru:\ (p,q\xTimes r) \to (r,\hat q \xTimes p\,,)
  \quad\;
  \frv:\ (p,q\xTimes r) \to (q,p \xTimes \hat r) \labl{5.2a}
by
  \be
  \fru(t) := (1_{\hat q} \xTimes t^*)\, \assoc^*_{\hat q,q,r}\, 
             (\eval^*_q \xTimes 1_r)\, \rho_r\,
			 \sqrt{d_r \over d_p}\,,\quad
  \frv(t) := (t^* \xTimes 1_{\hat r} )\, \assoc^*_{q,r,\hat r}\, 
             (1_{q} \xTimes \coev_r)\, \lambda_r\,
			 \sqrt{d_q \over d_p} \  \labl{5.2b}
and call them the Frobenius maps.
These maps are antilinear isometries, i.e.\ for the scalar product the formula
  \be
  (\fru(t_1),\fru(t_2)) = (t_2,t_1) = (\frv(t_1),\frv(t_2))\labl{5.3}
holds due to the trace property \erf{2.5}. They generate a $\ZZ_2$-graded 
antilinear and projective action of $S_3$ on the basic intertwiner spaces. A 
$\ZZ_2$-graded antilinear action 
of a $\ZZ_2$-graded group means linear (antilinear) action for even  
(odd) elements of the group. A permutation group is naturally $\ZZ_2$-graded
by distinguishing even and odd permutations. Moreover, the Frobenius maps 
generate only a projective action since using \erf{4.4} one proves that
  \be
  \fru^2 = \chi_q\cdot {id}_{(p,q\xtimes r)}\,,\qquad
  \frv^2 = \chi_r\cdot {id}_{(p,q\xtimes r)} \labl{5.4a}
as well as
  \be
  \fru\frv\fru = \frv\fru\frv 
  \colon\;\ (p,q\xTimes r)\to (\hat p,\hat r\xTimes\hat q)\,,\labl{5.4b}
which constitutes just a projective version of the defining
$S_3$-relations for the transpositions:
$\sigma_{12} \leftrightarrow \fru,\ \sigma_{23} \leftrightarrow \frv$.
In a generic case the Frobenius maps lead to an orbit of six
different intertwiner spaces
  \be \bearll
  \Im (id)   \!\!&= (p,q \xTimes r)\,,\nxl8
  \Im (\fru) \!\!&= (r,\hat q \xTimes p)\,,\nxl8
  \Im (\frv) \!\!&= (q,p \xTimes \hat r), \eear \qquad\bearll
  \Im (\fru\frv) = (\hat r,\hat p \xTimes q)\,,\nxl8
  \Im (\frv\fru) = (\hat q,r \xTimes \hat p)\,,\nxl8
  \Im (\frv\fru\frv) = (\hat p,\hat r \xTimes \hat q)\,. \eear \labl{5.5} 
It is easy to give the action of the Frobenius maps on canonical basic 
intertwiners $\rho_p\iN (p,\munit\xTimes p)$ and
$\lambda_p\iN (p,p\xTimes \munit)$, namely,
  \be
  \fru(\rho_p) = \rho_p\,, \qquad
  \frv(\rho_p) = {1 \over \sqrt{d_p}}\, \coev_p\,, \qquad
  \fru\frv(\rho_p) = \lambda_{\hat p} \labl{5.6a}
and
  \be
  \fru(\lambda_p) = {1 \over \sqrt{d_p}}\, \eval^*_p\,,  \qquad
  \frv(\lambda_p) =  \lambda_p\,, \qquad
  \frv\fru(\lambda_p) = \rho_{\hat p}\,.\labl{5.6b}
Therefore they are not generic but length three orbits of $S_3$
with a possible projective $\ZZ_2$-extension by $\chi_p$.

There are other cases, too, where the orbits are not generic.
Then we have a representation of the corresponding non-trivial stabilizer
subgroup of $S_3$ on a basic intertwiner space.  
In the list below we present all the possible irreducible representations of 
the stabilizer subgroups in the non-generic cases. In two cases we obtain 
certain restrictions on fusion coefficients.  
\vskip.3em

\noindent 
{\bf 1.}\, $q\eq\hat q$, $r\eq p\nE q$\,:\, orbit = 
$\{(p,p\xTimes q), (q,\hat p\xTimes p), (\hat p,q\xTimes \hat p)\}$\,.
\nxl9
Therefore there is a $\ZZ_2$-graded antilinear (projective if $\chi_q\eq{-}1$)
representation of $\ZZ_2$ on $(p,p\xTimes q)$
generated by $\frv$. But an antilinear unitary is always of order 2, 
therefore this intertwiner space is the null space if $\chi_q\eq{-}1$, i.e.\ the
corresponding fusion coefficient is $N^p_{pq}=0$. If $\chi_q\eq1$ we have a 
$\ZZ_2$-graded antilinear representation of $\ZZ_2$, and such \irrep s are 
unique up to unitary equivalence.
\nxl7
{\bf 2.}\, $q\eq r\eq \hat p\nE p$\,:\, orbit = 
$\{ (p,\hat p\xTimes \hat p), (\hat p,p\xTimes p)\}$\,.
\nxl9
Therefore there is a linear representation of $\ZZ_3$ on 
$(p,\hat p\xTimes \hat p)$, generated by $\fru\frv$.
One can use an orthonormal basis in 
$(p,\hat p\xTimes \hat p)$ in which the basis elements carry one
of the three possible (one-dimensional) \irrep s of $\ZZ_3$.
(A simple example is given in the Appendix.)
\nxl7
{\bf 3.}\, $q\eq r\eq p\eq\hat p$\,:\, orbit = $\{(p,p\xtimes p)\}$\,.
\nxl9
Therefore there is a $\ZZ_2$-graded antilinear (projective if $\chi_p\eq{-}1$) 
representation of $S_3$ on $(p,p\xTimes p)$.
Similarly to the first case we have $N^p_{pp}=0$ if $\chi_p\eq{-}1$. 
If $\chi_p\eq1$ we get a $\ZZ_2$-graded antilinear proper representation of 
$S_3$. Such \irrep s are one-dimensional and can be labelled by 
the three possible \irrep s of $\ZZ_3$. One can use a basis in 
$(p,p\xTimes p)$ where basis elements carry these \irrep s.

\SECTION{$S_4$-transformed simplicial maps of a tetrahedron }

Let us consider the following Hilbert space containing certain four-fold
tensor products of basic intertwiner spaces as orthogonal subspaces:
  \be \bearll
  \H\equiv \displaystyle\bigoplus_{\bf s} \H_{\bf s} \!
  &:= \displaystyle\bigoplus_{p,q,r,t,u,v\in\I}
  (p \xTimes q,u) \otimes   (u \xTimes r,t) \otimes
    (v,q \xTimes r) \otimes   (t,p \xTimes v) \nxl7
    &\;\equiv \displaystyle\bigoplus_{p,q,r,t,u,v\in\I}
  (u,p \xTimes q)^* \otimes   (t,u \xTimes r)^* \otimes
    (v,q \xTimes r) \otimes   (t,p \xTimes v) \,, \eear \labl{6.1}
where we put 
$(u,p \xTimes q)^*:=\{ t^*\,{\mid}\, t\iN(u,p \xTimes q)\}$.
Vectors in subspaces $\H_{\bf s}$ of $\H$ that correspond to a four-fold 
tensor product of basic intertwiner spaces can be labelled by the 
two-dimensional simplicial complex of the boundary of
a tetrahedron $\triangle^{\!3}$. More precisely, there are simplicial maps
  \be  \nu=(\nu^0,\nu^1,\nu^2):\ \partial\triangle^{\!3} \to \H_{\bf s} \ee
as follows.
$\nu^0$ maps all vertices into one point thus the vertices of
the tetrahedron remain unlabelled. (Nontrivial $\nu^0$ is
needed only in 2-categories which is out of the scope of the
present paper.) $\nu^1$ associates to each oriented edge
of $\triangle^{\!3}$ an irreducible object from the set of
representants $\I$. Thus the edges become labelled with $\I$:
  \be \bearll
  \nu^1(12):=p \,, &
  \nu^1(14):=t \,, \nxl6
  \nu^1(23):=q \,, &
  \nu^1(13):=u \,, \nxl6
  \nu^1(34):=r\quad \,, &
  \nu^1(24):=v \,. \eear  \labl{6.2}
Edges with opposite orientation carry the conjugate objects,
e.g., $  \nu^1(21)=\hat p$.

To every element of a basic intertwiner space one can associate
an oriented face by a right hand rule:\,%
\footnote{~In the vertex pictures, all arrows point downwards;
correspondingly we can safely suppress them.}

\hbox{
$(u,p \xTimes q) \ni \ \alpha \ = $ \hskip .5cm
\vbox{
\begin{xy} <2.3mm,0mm>:
  (0,2)   ,{ \Tu{ 3}{a}{\alpha} }, "au";"au"+(0,1)**@{-},
  "au"*+!DL{u},"ar"*+!DL{q},"al"*+!DR{p},
\end{xy}  }
\hskip 1cm $\longleftrightarrow$ \hskip .5cm
\vbox{\xy
0;<-\hrt cm,-.5cm>:<\hrt cm,-.5cm>::
(-1.5,-.5)="1", (.5,.5)="2", (-.5,-1.5)="3",
"1";"2" **@{-} ? *\dir{>} *+!UL{q},
"1";"3" **@{-} ? *\dir{>} *+!CD{u},
"3";"2" **@{-} ? *\dir{<} *+!UR{p},
{(-.4,-.4)*+{\alpha}},
\endxy}
\quad \vbox{ $\hskip -.5cm = \ \nu^2(123)_+$}
}
\vskip 1cm
\hbox{
$(u,p \xTimes q)^* \ni \ \alpha^* \ = $ \hskip .3cm
\vbox{
\begin{xy} <2.3mm,0mm>:
   (0,1.5)   ,{ \Tdr{ 3}{a}{\alpha^*} }, "ad";"ad"-(0,1)**@{-},
  "ad"*+!UL{u},"ar"*+!UL{q},"al"*+!UR{p},
\end{xy}  }
\hskip 1cm $\longleftrightarrow$ \hskip .5cm
\vbox{\xy
0;<\hrt cm,.5cm>:<-\hrt cm,.5cm>::
(-1.5,-.5)="1", (.5,.5)="2", (-.5,-1.5)="3",
"1";"2" **@{-} ? *\dir{<} *+!RD{p},
"1";"3" **@{-} ? *\dir{>} *+!CU{u},
"3";"2" **@{-} ? *\dir{>} *+!LD{q},
{(-.4,-.4)*+{\alpha^*}},
\endxy}
\quad\vbox{ $\hskip -.5cm = \ \nu^2(123)_-$}
}

\noindent
The two possible orientations tell us whether we are in a basic
intertwiner space $(u,p\xTimes q)$ or in its adjoint 
$(u,p\xTimes q)^*\equiv(p\xTimes q,u)$. Now the images of the faces of the
tetrahedron are defined to be elements of basic intertwiner spaces:
  \be \bearl
  \nu^2(123) :=\alpha_{pq}^u\,\in(u,p\xTimes q)
     \equiv(\nu^1(13),\nu^1(12)\xTimes \nu^1(23))\,, \nxl7
  \nu^2(134) := \beta_{ur}^t\,\in(t,u\xTimes r)
     \equiv(\nu^1(14),\nu^1(13)\xTimes \nu^1(34))\,, \nxl7
  \nu^2(234) := \gamma_{qr}^v\,\in(v,q\xTimes r)
     \equiv(\nu^1(24),\nu^1(23)\xTimes \nu^1(34))\,, \nxl7
  \nu^2(124) := \delta_{pv}^t\,\in(t,p\xTimes v)
     \equiv(\nu^1(14),\nu^1(12)\xTimes \nu^1(24))\,. \eear \ee

Permutation of the index set $\{1,2,3,4\}$ of the points of the
tetrahedron defines, using the Frobenius maps $\fru,\frv$, 
an $S_4$-action on the maps $\nu\colon\; \partial\triangle^{\!3}\to\H_{\bf s}$ 
as follows: 
  \be
  (\nu_{\tau})^0 := \nu^0 \circ \tau\,, \qquad
  (\nu_{\tau})^1(ij) := \nu^1(\tau(i),\tau(j))
  \qquad\mbox{for all }\; \tau\iN S_4 \;\ {\rm and}\;\ i,j\iN\{1,2,3,4\} \,.
  \labl{6.3}
The definition of the transforms of $\nu^2$ uses the
Frobenius maps. Let $\tau_{12},\tau_{23},\tau_{34}$
denote the generating transpositions of $S_4$. Then
  \be \bearl
  (\nu_{\tau_{12}})^2(123) := \fru( \nu^2(123)) \,, \nxl8
  (\nu_{\tau_{12}})^2(134) := \nu^2(234)\,,\eear \qquad\bearl
  (\nu_{\tau_{12}})^2(234) := \nu^2(134) \,, \nxl8
  (\nu_{\tau_{12}})^2(124) := \fru( \nu^2(124))\,; \eear \labl{6.4}
in other words, introducing the notation
  \be
  \alpha^*\oT\beta^*\oT\gamma\oT\delta =
  {\nu^2(123)}^*\oT {\nu^2(134)}^* \oT
  {\nu^2(234)} \oT  {\nu^2(124)}\in \H_{\bf s}\,,\labl{6.5a}
one has the induced transformations 
  \be \bearll
  \tau_{12}:& \alpha^*\oT\beta^*\oT\gamma\oT\delta 
  \to  \fru(\alpha)^*\oT\gamma^*\oT\beta\oT\fru(\delta)\,,\nxl7
  \tau_{23}:& \alpha^*\oT\beta^*\oT\gamma\oT\delta 
  \to  \frv(\alpha)^*\oT\delta^*\oT\fru(\gamma)\oT\beta\,,\nxl7
  \tau_{34}:& \alpha^*\oT\beta^*\oT\gamma\oT\delta 
  \to  \delta^*\oT\frv(\beta)^*\oT\frv(\gamma)\oT\alpha
  \eear \labl{6.5b} 
on the $\nu^2$-images.
The images of the original and the transformed
maps $\nu,\nu_{\tau_{12}},\nu_{\tau_{23}},\nu_{\tau_{34}}$
can be pictured as

$$\begin{array}{llll}
id: &
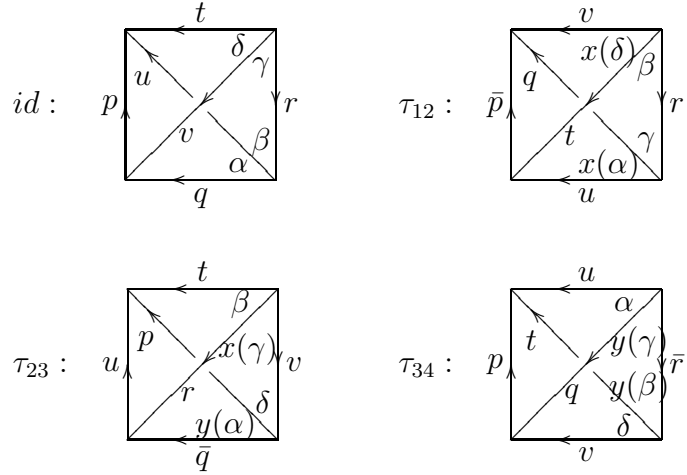
\begin{xy}
<10mm,0mm>: (.1,-.1)="+0",(-.1,.1)="-0",
(1,1)="1"    
,(1,-1)="2"  
,(-1,-1)="3" 
,(-1,1)="4"  
,
"1"-(.5,.2)*+{\delta},"1"-(.2,.5)*+{\gamma},
"2"+(-.5,.2)*+{\alpha},"2"+(-.2,.5)*+{\beta},
"1";"2"**@{-} ? *\dir{>} *+!LC{r},
"2";"3"**@{-} ?(.7) *\dir{>} ?*+!UC{q},
"3";"4"**@{-} ? *\dir{>} *+!RC{p},
"1";"4"**@{-} ?(.7) *\dir{>} ?*+!DC{t},
"1";"3"**@{-} ? *\dir{>} ?(0.7)*+!LC{v},
"2";"+0"**@{-},
"-0";"4"**@{-} ?(.7) *\dir{>} ?*+!RU{u}
\end{xy}
\qquad& \tau_{12}:&
\begin{xy}
<10mm,0mm>: (.1,-.1)="+0",(-.1,.1)="-0",
(1,1)="1"    
,(1,-1)="2"  
,(-1,-1)="3" 
,(-1,1)="4"  
,
"1"-(.7,.3)*+{x(\delta)},"1"-(.2,.5)*+{\beta},
"2"+(-.7,.2)*+{x(\alpha)},"2"+(-.2,.5)*+{\gamma},
"1";"2"**@{-} ? *\dir{>} *+!LC{r},
"2";"3"**@{-} ?(.7) *\dir{>} ?*+!UC{u},
"3";"4"**@{-} ? *\dir{>} *+!RC{\bar p},
"1";"4"**@{-} ?(.7) *\dir{>} ?*+!DC{v},
"1";"3"**@{-} ? *\dir{>} ?(0.7)*+!LC{t},
"2";"+0"**@{-},
"-0";"4"**@{-} ?(.7) *\dir{>} ?*+!RU{q}
\end{xy}
\\{}\\
\tau_{23}:&
\begin{xy}
<10mm,0mm>: (.1,-.1)="+0",(-.1,.1)="-0",
(1,1)="1"    
,(1,-1)="2"  
,(-1,-1)="3" 
,(-1,1)="4"  
,
"1"-(.5,.2)*+{\beta},"1"-(.4,.8)*+{x(\gamma)},
"2"+(-.7,.2)*+{y(\alpha)},"2"+(-.2,.5)*+{\delta},
"1";"2"**@{-} ? *\dir{>} *+!LC{v},
"2";"3"**@{-} ?(.7) *\dir{>} ?*+!UC{\bar q},
"3";"4"**@{-} ? *\dir{>} *+!RC{u},
"1";"4"**@{-} ?(.7) *\dir{>} ?*+!DC{t},
"1";"3"**@{-} ? *\dir{>} ?(0.7)*+!LC{r},
"2";"+0"**@{-},
"-0";"4"**@{-} ?(.7) *\dir{>} ?*+!RU{p}
\end{xy}
\qquad& \tau_{34}:&
\begin{xy}
<10mm,0mm>: (.1,-.1)="+0",(-.1,.1)="-0",
(1,1)="1"    
,(1,-1)="2"  
,(-1,-1)="3" 
,(-1,1)="4"  
,
"1"-(.5,.2)*+{\alpha},"1"-(.3,.7)*+{y(\gamma)},
"2"+(-.5,.2)*+{\delta},"2"+(-.3,.7)*+{y(\beta)},
"1";"2"**@{-} ?(.55) *\dir{>} ? *+!LC{\bar r},
"2";"3"**@{-} ?(.7) *\dir{>} ?*+!UC{v},
"3";"4"**@{-} ? *\dir{>} *+!RC{p},
"1";"4"**@{-} ?(.7) *\dir{>} ?*+!DC{u},
"1";"3"**@{-} ? *\dir{>} ?(0.7)*+!LC{q},
"2";"+0"**@{-},
"-0";"4"**@{-} ?(.7) *\dir{>} ?*+!RU{t}
\end{xy}
\end{array}$$
Extending the maps in \erf{6.5b} antilinearly, one obtains an action of $S_4$ on
the Hilbert space $\H$ in \erf{6.1}. The important property of this $S_4$ 
action is the following:

\begin{quote}
{\bf Lemma.} {\sl The above defined action of the 
transpositions $\tau_{12},\tau_{23},\tau_{34}$ induces a $\ZZ_2$-graded
antilinear representation of $S_4$ on $\H$.}
\end{quote}

\noindent{\it Proof.} \
One has to check only the defining relations of $S_4$:
  \be
  \tau_{12}^2=\tau_{23}^2=\tau_{34}^2= id\,, \qquad
  \tau_{12}\tau_{23}\tau_{12}= \tau_{23}\tau_{12}\tau_{23}\,, \qquad
  \tau_{23}\tau_{34}\tau_{23}= \tau_{34}\tau_{23}\tau_{34}\,.\labl{6.6}
Indeed one gets  
  \be \bearl
  \tau_{12}^2(\alpha^*\oT\beta^*\oT\gamma\oT\delta)= 
  \fru^2(\alpha)^*\oT\beta^*\oT\gamma\oT\fru^2(\delta)=
  \chi_p^2\cdot \alpha^*\oT\beta^*\oT\gamma\oT\delta=
  \alpha^*\oT\beta^*\oT\gamma\oT\delta\,, \nxl7
  \tau_{23}^2(\alpha^*\oT\beta^*\oT\gamma\oT\delta)= 
  \frv^2(\alpha)^*\oT\beta^*\oT\fru^2(\gamma)\oT\delta=
  \chi_q^2\cdot\alpha^*\oT\beta^*\oT\gamma\oT\delta=
  \alpha^*\oT\beta^*\oT\gamma\oT\delta\,,\nxl7
  \tau_{34}^2(\alpha^*\oT\beta^*\oT\gamma\oT\delta)= 
  \alpha^*\oT\frv^2(\beta)^*\oT\frv^2(\gamma)\oT\delta= 
  \chi_r^2\cdot\alpha^*\oT\beta^*\oT\gamma\oT\delta=
  \alpha^*\oT\beta^*\oT\gamma\oT\delta\,,
  \eear \labl{6.7a} 
while
  \be \bearll
  \tau_{12}\tau_{23}\tau_{12} (\alpha^*\oT\beta^*\oT\gamma\oT\delta) \!\!
  &=\fru\frv\fru(\alpha)^*\oT\fru(\beta)^* \oT\fru(\delta) \oT\fru(\gamma)\nxl8
  &=\frv\fru\frv(\alpha)^*\oT\fru(\beta)^*\oT \fru(\delta)\oT\fru(\gamma)
   =\tau_{23}\tau_{12}\tau_{23}(\alpha^* \oT\beta^*\oT\gamma\oT\delta)\,, \nxl6
  \tau_{23}\tau_{34}\tau_{23}(\alpha^*\oT\beta^*\oT\gamma\oT\delta) \!\!
  &=\frv(\beta)^*\oT\frv(\alpha)^*\oT\fru\frv\fru(\gamma)\oT\frv(\delta)\nxl8
  &=\frv(\beta)^*\oT\frv(\alpha)^*\oT\frv\fru\frv(\gamma)\oT\frv(\delta)
   =\tau_{34}\tau_{23}\tau_{34}(\alpha^*\oT\beta^*\oT\gamma\oT\delta)\,.
  \qed \eear \labl{6.7b}
Notice that although the Frobenius maps lead to a
($\ZZ_2$-graded antilinear) {\em projective\/} representation
of $S_3$ on each basic intertwiner space, due to the fact
that $\partial\triangle^{\!3}$ is a closed orientable surface
without boundary, the action of $S_4$ on ${\cal H}$ is a
{\em proper\/} (i.e., non-projective) representation. As a
matter of fact, according to \erf{6.7a}, the Frobenius
transformations $x^2$ and $y^2$ that lead to the signs
$\chi$ are always coming in pairs.

\SECTION{An $S_4$-invariant linear functional on $\H$}

Let us define a linear functional $\Phi\colon\; \H\to\CC$ on
the Hilbert space $\H$ \erf{6.1} by
  \be  \bearl
  \Phi(\alpha_{pq}^{u*}\ot \beta_{ur}^{t*}\ot \gamma_{qr}^v\ot \delta_{pv}^t)
  \nxl7 \qquad :=\sqrt{d_pd_qd_r\over d_t}\cdot
  \eval_t(1_{\hat t}\xTimes \beta_{ur}^{t*})(1_{\hat t}
  \xTimes (\alpha_{pq}^{u*}))
  (1_{\hat t}\xTimes\varphi_{p,q,r})
  (1_{\hat t}\xTimes(1_p\xTimes\gamma_{qr}^v))
  (1_{\hat t}\xTimes\delta_{pv}^t)\eval_t^* \,.  \eear \labl{7.1}
\def\Cr#1#2{\save\POS,c="#2r";
p-(#1,0)-(#1,0)="#2l",{\ellipse^{}} \restore }
\def\Er#1#2{\save\POS,c="#2r";
p-(#1,0)-(#1,0)="#2l",{\ellipse_{}} \restore }
\def\Cl#1#2{\save\POS,c="#2l";
p+(#1,0)+(#1,0)="#2r",{\ellipse_{}} \restore }
\def\El#1#2{\save\POS,c="#2l";
p+(#1,0)+(#1,0)="#2r",{\ellipse^{}} \restore }
Pictorially, the value is:
  \be \raisebox{.1em}{$
  \Phi(\alpha_{pq}^{u*}\ot \beta_{ur}^{t*}\ot \gamma_{qr}^v\ot \delta_{pv}^t)
  =\sqrt{d_pd_qd_r\over d_t}\,\cdot $} 
\raisebox{5.7em}{
\begin{xy} <2mm,0mm>:
   0,{ \Tu{3}{a}{\delta} },
   "ar",{ \Tu{2}{b}{\gamma}  },
   "bl";"bl"-(0,2)**@{-},{ \Tdr{2}{c}{\alpha^*}},
   "cd",{ \Tdl{3}{d}{\beta^*}},
   "au",{ \Cr{3}{C}},"dd",{ \Er{4}{E}},
   "al";"cl"**@{-},"br";"dr"**@{-},"Cl";"El"**@{-},
   "al"*+!CR{p},"ar"*+!LD{v},"au"*+!LD{t},
   "cr"*+!LU{q},"dr"*+!LD{r},"dl"*+!RU{u},
\end{xy} }
  \ee

\noindent
(In order not to overburden the picture, we do not indicate 
the maps $\lambda, \rho, \varphi$; due to coherence
and naturality they can be put back unambiguously.)

The action of $\tau\iN S_4$ on $\Phi$ is the one induced by the action 
on $\cal H$,
  \be 
  (\tau\Phi)(\alpha^*\ot\beta^*\ot\gamma\ot\delta):=
  \overline{\Phi(\tau^{-1}(\alpha^*\ot\beta^*\ot\gamma\ot
  \delta))}^{\deg\tau}\,,\labl{7.2}
where the notation ``overline to the power of $\deg\,\tau$" means complex
conjugation for odd $S_4$-elements and the identity operation
for even elements. This transformation property is required by 
the linearity of $\Phi$ and the antilinearity of $S_4$ on $\H$ to be compatible:
  \be 
  (\tau\Phi)(\lambda X)=
  \overline{\Phi(\tau^{-1}(\lambda X))}^{\deg\tau}=
  \overline{{\bar\lambda}^{\deg\tau}_{\phantom|}\Phi(\tau^{-1}(X))}^{\deg \tau}=
  \lambda\cdot\overline{\Phi(\tau^{-1}(X))}^{\deg \tau}\labl{7.3}
for $X\iN \H$. Hence the functional $\Phi$ is $S_4$-invariant if 
  \be
  (\tau\Phi)(X):=\overline{\Phi(\tau^{-1}(X))}^{\deg \tau}=\Phi(X)\labl{7.4a}
for all $\tau\iN S_4$, or equivalently
  \be 
  \Phi(\tau^{-1}(X))=\overline{\Phi(X)}^{\deg \tau}\labl{7.4b}
for all $\tau\iN S_4$.

\begin{quote}
{\bf Proposition.} {\sl The functional $\Phi$ is constant on $S_4$-orbits.}
\end{quote}

\noindent{\it Proof.} It is sufficient to show this property for the
generators of $S_4$. Below there is a diagrammatic proof where we
used the definitions of the Frobenius maps, the trace property, 
rigidity, the property ${\rm End}\, \munit=\CC\cdot 1_\munit$ and 
sphericity in the last series of pictures.\,%
\footnote{~The figures are collected on separate pages at the end of the file.
A version of the paper where the figures are directly included in the equations
can be downloaded from {\small\tt 
http:/$\!$/www.desy.de/\raisebox{-.2em}{$\tilde{\phantom.}$}jfuchs/s4/s4.ps.gz}.
}

Invariance with respect to $\tau_{12}$ is shown by
the following chain of equalities:
$$\begin{array}{ccccc}
  \leavevmode \sfbox{4}{1}
& \raisebox{.4em}{$\quad \stackrel{1}{=} \quad $} 
& \leavevmode \sfbox{5}{2}
& \raisebox{.4em}{$\quad \stackrel{2}{=} \quad $} 
& \leavevmode \sfbox{6}{3}
\nxl5
& \raisebox{.4em}{$\quad \stackrel{3}{=} \quad $} 
& \leavevmode \sfbox{7}{4}
& \raisebox{.4em}{$\quad \stackrel{4}{=} \quad $} 
& \leavevmode \sfbox{8}{5}
\eear$$
Here we have used
the definition of the Frobenius map $x$ in the first equality,
the monoidality of the functor of taking duals in the second,
used the trace property in the third, and
the definition of `$\,{}^*\,$' in the fourth equality.

\noindent
Proof of the invariance with respect to $\tau_{23}$:
$$\begin{array}{ccccc}
  \leavevmode \sfbox{16}{6}
& \raisebox{.4em}{$\quad \stackrel{1}{=} \quad $}
& \leavevmode \sfbox{17}{7}
\nxl5
& \raisebox{.4em}{$\quad \stackrel{2}{=} \quad $}
& \leavevmode \sfbox{19}{8}
& \raisebox{.4em}{$\quad \stackrel{3}{=} \quad $}
& \leavevmode \sfbox{20}{9}
\eear$$   
This chain of equalities is obtained as follows. In the first
equality we used the definition of the Frobenius maps $x$ and $y$;
in the second equality the rigidity identity is used; and in
the third equality the definition of `$\,{}^*\,$' is implemented.

\noindent
Proof of the invariance with respect to $\tau_{34}$:
$$\begin{array}{ccccc}
  \leavevmode \sfbox{9}{10}
& \raisebox{.4em}{$\quad \stackrel{1}{=} \quad $} 
& \leavevmode \sfbox{10}{11}
& \raisebox{.4em}{$\quad \stackrel{2}{=} \quad $} 
& \leavevmode \sfbox{11}{12}
\nxl5
& \raisebox{.4em}{$\quad \stackrel{3}{=} \quad $} 
& \leavevmode \sfbox{12}{13}
& \raisebox{.4em}{$\quad \stackrel{4}{=} \quad $} 
& \leavevmode \sfbox{13}{14}
\nxl5
& \raisebox{.4em}{$\quad \stackrel{5}{=} \quad $} 
& \leavevmode \sfbox{14}{15}
& \raisebox{.4em}{$\quad \stackrel{6}{=} \quad $} 
& \leavevmode \sfbox{15}{16}
\eear$$
In this chain of equalities we have used
the definition of the Frobenius map $y$ in the first
equality; the spherical property in the second;
the monoidality of the functor of taking duals in the
third; used the trace property in the fourth;
used the spherical property in the fifth;
and used the definition of `$\,{}^*\,$' in the sixth.
\qed

By definition the normalized $F$-coefficients are 
  \be
  \Hat F^{(pqr)_t}_{u,v}(\alpha^*\ot\beta^*\ot\gamma\ot\delta):=
  \Phi(\alpha_{pq}^{u*}\ot \beta_{ur}^{t*}\ot\gamma_{qr}^v\ot
  \delta_{pv}^t)\,,\labl{7.6}
where $\alpha,\beta,\gamma,\delta$ are elements of orthonormal
bases of the corresponding basic intertwiner spaces. Their relation to the 
$6j$-symbols $\{ F^{(pqr)_t}_{\alpha u\beta,\gamma v\delta}\}$ is given by 
  \be
  F^{(pqr)_t}_{\alpha u\beta,\gamma v\delta}
  ={1\over \sqrt{d_pd_qd_rd_t}}\,
  \Hat F^{(pqr)_t}_{u,v}(\alpha^*\oT\beta^*\oT\gamma\oT\delta) \,.  \ee
Because of the isometry property of $\varphi$ they are unitary matrices in 
the multi-labels $(\alpha u\beta, \gamma v\delta)$.
The proposition leads to the possibility of computing $\Hat F$ in a single
point of an $S_4$-orbit and determine the $6j$-symbols on all the
other elements of that orbit. 

\newpage
\appendix
\SECTION{Appendix}

As a simple but nevertheless non-trivial illustration of the results of
the main text let us consider the following three (degenerate) rational 
Hopf algebras \cite{vecs,fugv3} that can be obtained as deformations of 
the Hopf algebra $\CC\ZZ_3$, i.e.\ of the group algebra of the cyclic 
group $\ZZ_3$. The structural data can be summarized as follows.
  \be
  H=\CC e_0\oplus \CC e_1\oplus \CC e_2\,,\qquad 
  e_p^*=e_p^2=e_p\;\ {\rm for}\;\ i\eq 0,1,2\,,\labl{A.1a}
  \be
  \Delta(e_p)=\sum_{q,r=0\atop q+r=p\,{\rm mod}\, 3}^2 e_q\oT e_r\,,\qquad
  S(e_p)=e_{-p\,{\rm mod}\, 3}\,,\labl{A.1b}
  \be
  \lambda=\rho={\bf 1}\equiv e_0+e_1+e_2=l=r\in H\,,\labl{A.1c}
  \be
   \varphi=\sum_{p,q,r=0}^2\omega_{pqr}\cdot e_p\oT e_q\oT e_r\in
   H\otimes H\otimes H \,. \labl{A.1d}
Here $\omega_{111}\eq\omega_{222}\eq\omega_{112}\eq\omega_{221}\eq\omega_{211}
\eq\omega_{122}=:\omega$ is a third root of unity, $\omega^3=1$, which
parametrizes the three different rational Hopf algebras, while in all 
three cases one has $\omega_{pqr}=1$ for all other combinations of indices.

The representation category ${\bf Rep}\, H$ is a rigid monoidal 
$C^*$-category. The \irrep s $D_p,\, p=0,1,2$ are one-dimensional and obey
$D_p(e_q)=\delta_{p,q}$. The basic intertwiner spaces are one-dimensional 
at most, and we can choose $(p,q\xTimes r)=\CC\cdot 1_{qr}^p$ for the
non-trivial ones, where $1_{qr}^p$ maps the tensor product of the chosen unit
vectors into the chosen unit vector of the corresponding one-dimensional 
representation spaces, i.e.\ $1_{qr}^p(v_q\oT v_r)=v_p$. The natural isometries
connected to monoidality and the standard rigidity intertwiners are given by
  \be
  \lambda_p=1_{p0}^p\,,\qquad\rho_p=1_{0p}^p\,,\qquad 
  \varphi_{p,q,r}=(D_p\,{\otimes}\, D_q\,{\otimes}\, D_r)(\varphi)\,,
  \ee
  \be
  e_p=1_{\hat pp}^{0*}\,,\qquad c_p=1_{p\hat p}^0\,.
  \ee
The values of $\chi$ \erf{4.4} are all trivial: $\chi_p\eq 1$ for $p\eq 0,1,2$. 
Using the definitions \erf{5.2b} of the Frobenius maps, one obtains 
  \be
  x(1_{qr}^p)=\bar\omega_{\hat qqr}\,1_{\hat qp}^r\,,\qquad
  y(1_{qr}^p)=\omega_{qr\hat r}\,1_{p\hat r}^q\,,
  \ee
with $\omega_{pqr}$ as in \erf{A.1d}.
In the case $(p,q\xTimes r)=(1,2\xTimes 2)$, this leads to 
  \be
  xy(1_{22}^1)=\omega\cdot 1_{22}^1\,,
  \ee
which serve as examples of a degenerate orbit of type 2 in section \ref{fmbi} 
with the possible third roots of unity.

The $S_4$-symmetry is valid in a non-trivial way in the following sense.
First note that there are five $S_4$-orbits of normalized $F$-coefficients
$\Hat F^{(pqr)_t}_{u,v}(\alpha^{u*}_{pq}\oT\beta^{t*}_{ur}\oT
\gamma^v_{qr}\oT\delta^t_{pv})$ that have the elements (only their
$pqr$ edges indicated):
  \be  \bearl
  \{ 000\}\,,\quad \{ 001,200,120,012\}\,,\quad \{ 002,100,210,021\}\,, \nxl7
  \{ 010,020,102,201,121,212\}\,, \nxl7
  \{ 101,011,110,202,022,220,111,222,112,122,221,211\}\,,
  \eear \ee
respectively, together with their complex conjugated quantities.
One can easily compute the first normalized $F$-coefficient of the fifth
orbit: 
  \be
  \Hat F^{(101)_2}_{1,1}(1^{1*}_{10}\oT 1^{2*}_{11}\oT 1^1_{01}\oT 1^2_{11})=1
  \,; \ee
then due to the $S_4$-invariance one deduces that the other coefficients in 
the same orbit have the value 1, too. Computation of the 
$\tau_{12}$-transformed quantity
  \be  \bearll  1 \!\!
  &=\tau_{12}(\Hat F^{(101)_2}_{1,1}
    (1^{1*}_{10}\ot 1^{2*}_{11}\ot 1^1_{01}\ot 1^2_{11}))
  =\overline{\Hat F^{(211)_1}_{0,2}
    (x(1^1_{10})^*\ot 1^{1*}_{01}\ot 1^2_{11}\ot x(1^2_{11}))}\nxl6
  &=\overline{\Hat F^{(211)_1}_{0,2}
    (\omega_{210}\bar\omega_{211}\cdot 
       1^{0*}_{21}\ot 1^{1*}_{01}\ot 1^2_{11}\ot 1^1_{22})}
  =\omega\cdot\overline{\Hat F^{(211)_1}_{0,2}
     (1^{0*}_{21}\ot 1^{1*}_{01}\ot 1^2_{11}\ot 1^1_{22})}
  =\omega\cdot\bar\omega \eear \ee
then shows that one may not have $S_4$-invariance for a fixed set of basic 
intertwiners in general, so it is important to take the action of the 
Frobenius maps into account. In this example even the following stronger 
statement holds: in the case of $\omega\not=1$ there is {\em no\/}
such choice for the set of 
orthonormal basic intertwiners, $\{ b_{qr}^{q\xtimes r}\,{:=}\,\omega_{qr}\cdot 
1_{qr}^{q\xTimes r}\,\vert\,q,r\eq 0,1,2\}$ with arbitrary but fixed values of 
$\omega_{qr}\iN\CC,\;\vert\omega_{qr}\vert\eq 1$, that leads to a constant 
value of normalized $F$-coefficients within the chosen set of intertwiners.
Indeed, the relation
  \be \bearll \Hat F^{(101)_2}_{1,1}
     (b^{1*}_{10}\ot b^{2*}_{11}\ot b^1_{01}\ot b^2_{11}) \!\!&=
  \tau_{13}(\Hat F^{(101)_2}_{1,1}
     (b^{1*}_{10}\ot b^{2*}_{11}\ot b^1_{01}\ot b^2_{11})) \nxl7&=
  \overline{\Hat F^{(101)_2}_{1,1}
     (y(b^1_{10})^*\ot b^{2*}_{11}\ot x(b^1_{01})\ot b^2_{11})}
  \eear \ee
implies that the common value should be $\pm 1$, but then the product
of the $\tau_{12}\tau_{34}$- and 
$\tau_{23}\tau_{34}\tau_{12}\tau_{23}$-transformed $F$-coefficients
leads to the contradiction
  \be  \bearll
  1 \not= \!& \Hat F^{(222)_0}_{1,1}
  (x(b^2_{11})^*\ot y(b^1_{01})^*\ot y(b^2_{11})\ot x(b^1_{10}))   \nxl7 &
  \cdot \Hat F^{(111)_0}_{2,2}
  (xy(b^2_{11})^*\ot xy(b^1_{01})^*\ot xy(b^2_{11})\ot yx(b^1_{10}))
  =\omega^2\,.  \eear \ee

We also note that for a non-trivial third root of unity, $\omega\not=1$, there 
are no solutions of the hexagon equations, that is ${\bf Rep}\, H$ cannot be
made into a braided category in those cases.

\medskip
\small
\noindent{\bf Acknowledgements:}\\
A.\ Ganchev acknowledges the support of the Alexander von Humboldt
Foundation and the hospitality of FB Physik, University of Kaiserlautern
and of II.\ Institut f\"ur Theoretische Physik, University of Hamburg.
The work of A.\ Ganchev, K.\ Szlach\'anyi and P. Vecserny\'es was supported 
in part by a joint project of the Bulgarian and Hungarian Academies of Science
and by the Hungarian Scientific Research Fund (OTKA T 020 895).

\newpage

 \def\wb{\,\linebreak[0]} \def\wB {$\,$\wb}
 \def\Bi{\bibitem }
 \newcommand\Erra[3]  {\,[{\em ibid.}\ {#1} ({#2}) {#3}, {\em Erratum}]}
 \newcommand\BOOK[4]  {{\em #1\/} ({#2}, {#3} {#4})}
 \newcommand\J[5]     {\ {\sl #5}, {#1} {#2} ({#3}) {#4} }
 \newcommand\Prep[2]  {{\sl #2}, preprint {#1}}
 \newcommand\talku[2] {{\sl #2}, talk presented at the {#1} (unpublished)}
 \newcommand\inBO[7]  {\ {\sl #7},
                      in:\ {\em #1}, {#2}\ ({#3}, {#4} {#5}), p.\ {#6}}
 \newcommand\iNBO[7]  {\ {\sl #7},
                      in:\ {\em #1}, {#2}\ ({#3}, {#4} {#5}) }
   \def\anop  {Ann.\wb Phys.}
   \newcommand\ilag[2] {\inBO{\Infdim Lie \A s and Groups {\rm[Adv.\
              Series in Math.\ Phys.\ 7]}} {V.G.\ Kac, ed.} \WS\Si{1989}
              {{#1}}{{#2}}}
   \def\comp  {Com\-mun.\wb Math.\wb Phys.}
   \def\ijmp  {Int.\wb J.\wb Mod.\wb Phys.\ A}
   \def\injm  {Int.\wb J.\wb Math.}
   \def\jopa  {J.\wb Phys.\ A}
   \def\jpaa  {J.\wB Pure\wB Appl.\wb Alg.}
   \def\lemp  {Lett.\wb Math.\wb Phys.}
   \def\npbp  {Nucl.\wb Phys.\ B (Proc.\wb Suppl.)}
   \def\nupb  {Nucl.\wb Phys.\ B}
   \def\phlb  {Phys.\wb Lett.\ B}
   \def\remp  {Rev.\wb Mod.\wb Phys.}
   \def\rvmp  {Rev.\wb Math.\wb Phys.}
   \def\tams  {Trans.\wb Amer.\wb Math.\wb Soc.}
   \def\topo  {Topology}
\def\A       {Algebra}
\def\alg     {algebra}
\def\AP      {{Academic Press}}
\def\Be      {{Berlin}}
\def\BC      {{Ben\-jamin\,/\,Cum\-mings}}
\def\dim     {dimension}
\def\fts     {field theories}
\def\jf      {J.\ Fuchs}
\def\hopf    {Hopf algebra}
\def\Infdim  {Infinite-dimensional }
\def\NY      {{New York}}
\def\q       {quantum }
\def\Q       {Quantum }
\def\qfts    {quantum field theories}
\def\qg      {quantum group}
\def\Rep     {Representation}
\def\Si      {{Singapore}}
\def\ssi     {semisimple}
\def\stc     {statistic}
\def\sym     {symmetry}
\def\tft     {topological field theory}
\def\tfts    {topological field theories}
\def\trfo    {transformation}
\def\SV      {{Sprin\-ger Verlag}}
\def\WS      {{World Scientific}}

 \sfpic{4}{1}

\mbox{
\begin{xy} <2.3mm,0mm>:
   0   ,{ \Tu{ 3}{a}{x(\delta)} },
   "ar",{ \Tu{ 2}{b}{\beta}  },
   "bl",{ \Tdr{2}{c}{x(\alpha)^*}},
   "cd",{ \Tdl{3}{d}{\gamma^*}},
   "au",{ \Cr{3}{C}},"dd",{ \Er{4}{E}},
   "al";"cl"**@{-}?(0.5)*+!CR{\bar p},
   "br";"dr"**@{-}?(0.5)*+!CL{r},
   "Cl";"El"**@{-},"dl"*+!UR{q},
   "au"*+!DL{v},"ar"*+!DL{t},"bl"*+!UL{u}
\end{xy} }

 \sfpic{5}{2}

\mbox{
\begin{xy} <2.5mm,0mm>:
   0   ,{ \Tdr{1.5}{d}{\delta^*} },
   "dd",{ \Tu{ 1.5}{b}{\beta}  },
   "bl",{ \Tu{ 1  }{a}{\alpha}},
   "ar",{ \Tdl{1  }{c}{\gamma^*}},
   "dl",{ \Cr{1.5}{cc}},"al",{ \Er{1}{ee}},
   "dr",{ \Cr{4.5}{C}},"cd",{ \Er{4}{E}},
   "br";"cr"**@{-}?(0.5)*+!CL{r},
   "ccl";"eel"**@{-}?(0.5)*+!RD{\bar p},
   "Cl";"El"**@{-},
   "bl"*+!RD{u},"dd"*+!LC{t},
   "cl"-(0,1)*+!RU{q},"dr"*+!LD{v}
\end{xy} }

 \sfpic{6}{3}

\mbox{
\begin{xy} <2.5mm,0mm>:
   0   ,{ \Tdr{1.5}{d}{\delta^*} },
   "dd",{ \Tu{ 1.5}{b}{\beta}  },
   "bl",{ \Tu{ 1  }{a}{\alpha}},
   "ar",{ \Tdl{1  }{c}{\gamma^*}},
   "dl",{ \Cr{2.5}{cc}},"al",{ \Er{2}{ee}},
   "dr",{ \Cr{4.5}{C}},"cd",{ \Er{4}{E}},
   "br";"cr"**@{-}?(0.5)*+!CL{r},
   "ccl";"eel"**@{-},"Cl";"El"**@{-},
   "bl"*+!RD{u},"dl"*+!RU{p},"dd"*+!LC{t},
   "cl"-(0,1)*+!RU{q},"dr"*+!LD{v}
\end{xy} }

 \sfpic{7}{4}

\mbox{
\begin{xy} <2mm,0mm>:
   0   ,{ \Tu{3}{a}{\beta} },
   "al",{ \Tu{2}{b}{\alpha}  },
   "br";"br"-(0,2)**@{-},{ \Tdl{2}{c}{\gamma^*}},
   "cd",{ \Tdr{3}{d}{\delta^*}},
   "au",{ \Cr{4}{C}},"dd",{ \Er{3}{E}},
   "ar";"cr"**@{-}?(0.5)*+!RU{p},
   "bl";"dl"**@{-},"Cl";"El"**@{-},
   "al"*+!RD{u},"ar"*+!LD{r},"au"*+!LD{t},
   "cl"*+!RU{q},"dr"*+!LU{v},
\end{xy} }

 \sfpic{8}{5}

\mbox{
\begin{xy} <2mm,0mm>:
   0,{ \Tu{3}{a}{\delta} },
   "ar",{ \Tu{2}{b}{\gamma}  },
   "bl";"bl"-(0,2)**@{-},{ \Tdr{2}{c}{\alpha^*}},
   "cd",{ \Tdl{3}{d}{\beta^*}},
   "au",{ \Cr{3}{C}},"dd",{ \Er{4}{E}},
   "al";"cl"**@{-},"br";"dr"**@{-},"Cl";"El"**@{-},
   "al"*+!CR{p},"ar"*+!LD{v},"au"*+!LD{t},
   "cr"*+!LU{q},"dr"*+!LD{r},"dl"*+!RU{u},
   (-5,4);(6,4)**@{-}
\end{xy} }

 \sfpic{16}{6}

\begin{xy}
<2.3mm,0mm>:
0,{ \Tu{3}{a}{\beta} },
"ar",{ \Tu{2}{b}{x(\gamma)}  },
"bl";"bl"-(0,2)**@{-},{ \Tdr{2}{c}{y(\alpha)^*}},
"cd",{ \Tdl{3}{d}{\delta^*}},
"au",{ \Cr{3}{C}},"dd",{ \Er{4}{E}},
"al";"cl"**@{-},"br";"dr"**@{-},"Cl";"El"**@{-},
"al"*+!CR{u},"ar"*+!LD{r},"au"*+!LD{t},
"cr"*+!LU{\bar q},"dr"*+!LD{v},"dl"*+!RU{p},
\end{xy}

 \sfpic{17}{7}

\begin{xy}
<2mm,0mm>:
0   ,{  \Tu{4.5}{b}{\beta} },
"br",{ \Tdr{1.64}{c}{\gamma^*}  },
"cl",{ \Cr{1}{cc}},
"bl";p-(0,5)**@{-},
     { \Tu{1.64}{a}{\alpha}},
"ar",{ \El{1}{ee}},
"al",{ \Tdl{4.5}{d}{\delta^*}},
"ccl";"eer"**@{-},"dr";"cd"**@{-},
"bu",{ \Cr{5}{C}},"dd",{ \Er{4.2}{E}},
"Cl";"El"**@{-},
"bu"*+!LD{t},"cr"*+!LD{r},
"dl"*+!RU{p},"bl"*+!RU{u},"dr"*+!LD{v},
"eer"*+!LD{\bar q},
\end{xy}

 \sfpic{19}{8}

\begin{xy}
<2mm,0mm>:
0   ,{ \Tu{3}{a}{\beta} },
"al",{ \Tu{2}{b}{\alpha}  },
"br";"br"-(0,2)**@{-},{ \Tdl{2}{c}{\gamma^*}},
"cd",{ \Tdr{3}{d}{\delta^*}},
"au",{ \Cr{4}{C}},"dd",{ \Er{3}{E}},
"ar";"cr"**@{-},"bl";"dl"**@{-},"Cl";"El"**@{-},
"al"*+!LD{u},"ar"*+!LD{r},"au"*+!LD{t},
"cl"*+!RU{q},"dr"*+!LU{v},"dl"*+!RU{p},
\end{xy}

 \sfpic{20}{9}

\begin{xy}
<2mm,0mm>: 0,{ \Tu{3}{a}{\delta} },
"ar",{ \Tu{2}{b}{\gamma}  },
"bl";"bl"-(0,2)**@{-},{ \Tdr{2}{c}{\alpha^*}},
"cd",{ \Tdl{3}{d}{\beta^*}},
"au",{ \Cr{3}{C}},"dd",{ \Er{4}{E}},
"al";"cl"**@{-},"br";"dr"**@{-},"Cl";"El"**@{-},
"al"*+!CR{p},"ar"*+!LD{v},"au"*+!LD{t},
"cr"*+!LU{q},"dr"*+!LD{r},"dl"*+!RU{u},
(-5,4);(6,4)**@{-}
\end{xy}

 \sfpic{9}{10}

\mbox{
\begin{xy} <2mm,0mm>:
0   ,{ \Tu{3}{a}{\alpha} },
"ar",{ \Tu{2}{b}{y(\gamma)}  },
"bl",{ \Tdr{2}{c}{\delta^*}},
"cd",{ \Tdl{3}{d}{y(\beta)^*}},
"au",{ \Cr{3}{C}},"dd",{ \Er{4}{E}},
"al";"cl"**@{-}?(0.5)*+!CR{p},
"br";"dr"**@{-}?(0.5)*+!CL{\bar r},
"Cl";"El"**@{-},
"dl"*+!UR{t},
"au"*+!DL{u},"ar"*+!DL{q},"bl"*+!DR{v}
\end{xy} }

 \sfpic{10}{11}

\mbox{
\begin{xy} <2mm,0mm>:
0 ,{  \Tu{2}{a}{\alpha} },
"ar",{ \Tdl{2}{c}{\gamma^*}  },
"cd",{ \Tdr{3}{d}{\delta^*}},
"dd",{  \Tu{3}{b}{\beta}},
"cr",{ \Cl{2}{cc}},"br",{ \El{3}{ee}},
"au",{ \Cr{4}{C}},"bl",{ \Er{3}{E}},
"al";"dl"**@{-}?(0.5)*+!CR{p},
"ccr";"eer"**@{-}?(0.5)*+!CL{\bar r},
"Cl";"El"**@{-},
"au"*+!DR{u},"ar"*+!UR{q},"bu"*+!CR{t},
"cd"-(0,1)*+!CL{v},
\end{xy} }

 \sfpic{11}{12}

\mbox{
\begin{xy} <2.5mm,0mm>:
0   ,{ \Tu{1}{a}{\alpha} },
"ar",{ \Tdl{1}{c}{\gamma^*}  },
"cd"
,{ \Tdr{1.5}{d}{\delta^*}},
"dd",{ \Tu{1.5}{b}{\beta}},
"cr",{ \Cl{1}{cc}},"br",{ \El{1.5}{ee}},
"au",{ \Cl{4}{C}},"bl",{ \El{4.5}{E}},
"al";"dl"**@{-}?(0.5)*+!CR{p},
"ccr";"eer"**@{-}?(0.5)*+!RD{\bar r},
"Cr";"Er"**@{-},
"ar"+(0,1)*+!LD{q},"au"*+!RD{u},
"cd"*+!LU{v},"dd"*+!RC{t}
\end{xy}}

 \sfpic{12}{13}

\mbox{
\begin{xy} <2.5mm,0mm>:
0   ,{ \Tu{1}{a}{\alpha} },
"ar",{ \Tdl{1}{c}{\gamma^*}  },
"cd"
,{ \Tdr{1.5}{d}{\delta^*}},
"dd",{ \Tu{1.5}{b}{\beta}},
"cr";"cr"+(0,2)**@{-},{ \Cl{2}{cc}},
"br";"br"+(0,1)**@{-},{ \El{2.5}{ee}},
"au",{ \Cl{4}{C}},"bl",{ \El{4.5}{E}},
"al";"dl"**@{-}?(0.5)*+!CR{p},
"ccr";"eer"**@{-}?(0.5)*+!RD{\bar r},
"Cr";"Er"**@{-},
"ar"+(0,1)*+!LD{q},"au"*+!RD{u},
"cd"*+!LU{v},"dd"*+!RC{t}
\end{xy} }

 \sfpic{13}{14}

\mbox{
\begin{xy} <2mm,0mm>:
0   ,{ \Tu{3}{a}{\beta} },
"al",{ \Tu{2}{b}{\alpha}  },
"br";"br"-(0,2)**@{-}?(0.5)*+!CR{q}
    ,{ \Tdl{2}{c}{\gamma^*}},
"cd",{ \Tdr{3}{d}{\delta^*}},
"au",{ \Cl{3}{C}},"dd",{ \El{4}{E}},
"bl";"dl"**@{-}?(0.5)*+!CR{p},
"ar";"cr"**@{-}?(0.5)*+!CR{r},
"Cr";"Er"**@{-},
"al"*+!DR{u},
"au"*+!RD{t},
"dr"*+!LU{v},
\end{xy} }

 \sfpic{14}{15}

\mbox{
\begin{xy} <2mm,0mm>:
0   ,{ \Tu{3}{a}{\beta} },
"al",{ \Tu{2}{b}{\alpha}  },
"br";"br"-(0,2)**@{-},{ \Tdl{2}{c}{\gamma^*}},
"cd",{ \Tdr{3}{d}{\delta^*}},
"au",{ \Cr{4}{C}},"dd",{ \Er{3}{E}},
"ar";"cr"**@{-}?(0.5)*+!RU{p},
"bl";"dl"**@{-},"Cl";"El"**@{-},
"al"*+!RD{u},"ar"*+!LD{r},"au"*+!LD{t},
"cl"*+!RU{q},"dr"*+!LU{v},
\end{xy} }

 \sfpic{15}{16}

\mbox{
\begin{xy} <2mm,0mm>:
0,{ \Tu{3}{a}{\delta} },
"ar",{ \Tu{2}{b}{\gamma}  },
"bl";"bl"-(0,2)**@{-},{ \Tdr{2}{c}{\alpha^*}},
"cd",{ \Tdl{3}{d}{\beta^*}},
"au",{ \Cr{3}{C}},"dd",{ \Er{4}{E}},
"al";"cl"**@{-},"br";"dr"**@{-},"Cl";"El"**@{-},
"al"*+!CR{p},"ar"*+!LD{v},"au"*+!LD{t},
"cr"*+!LU{q},"dr"*+!LD{r},"dl"*+!RU{u},
(-5,4);(6,4)**@{-}
\end{xy} }

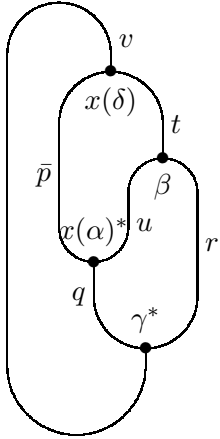
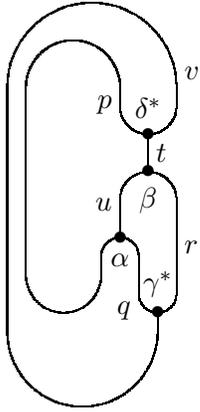
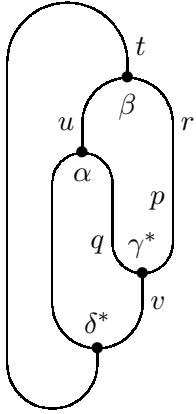
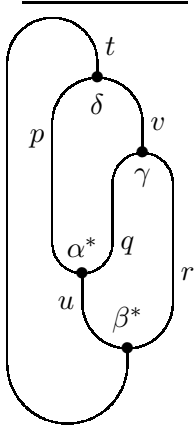
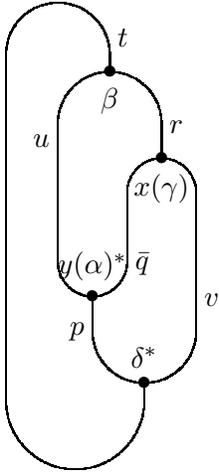
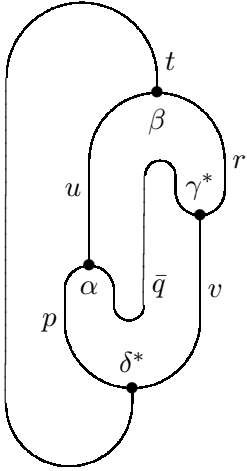
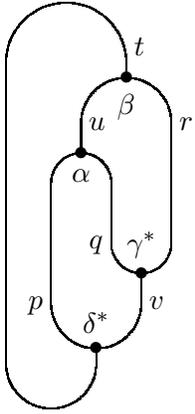
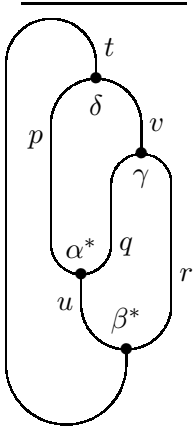
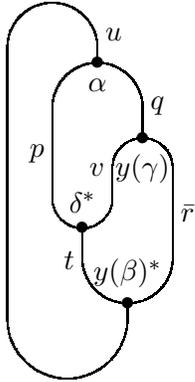
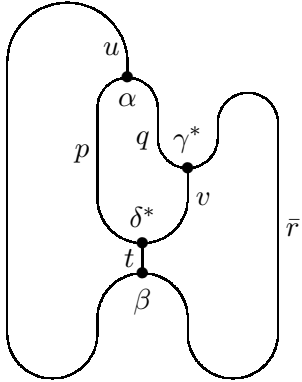
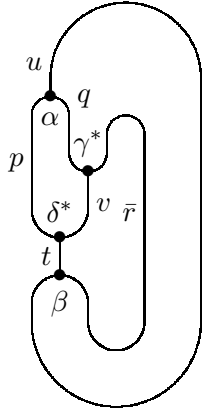
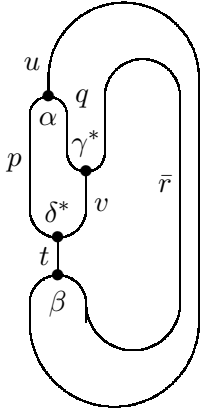
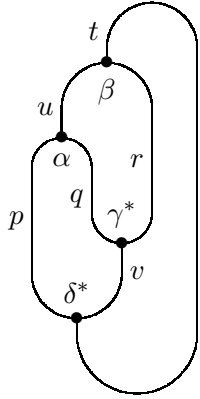
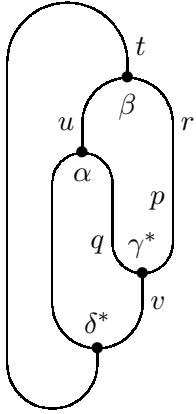
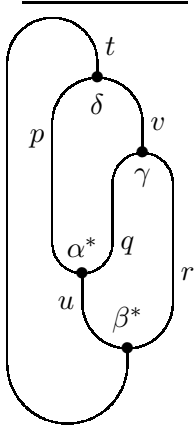
\end{document}